\documentclass[10pt,conference]{IEEEtran}

\usepackage{cite}
\usepackage{amsmath,amssymb,amsfonts}
\usepackage{graphicx}
\usepackage{multirow}
\usepackage{textcomp}
\usepackage{xcolor}
\usepackage{hyperref}

\usepackage{soul}
\usepackage{makecell}

\usepackage{tabulary}
\usepackage{subcaption}

\hypersetup
{
  pdftitle   = {import2vec: learning embeddings for software libraries},
  pdfsubject = {import2vec: learning embeddings for software libraries},
}

\usepackage{listings}
\begin{document}

\title{Import2vec\\Learning Embeddings for Software Libraries}


\author{\IEEEauthorblockN{Bart Theeten}
\IEEEauthorblockA{
\textit{Nokia Bell Labs}\\
Antwerp, Belgium \\
bart.theeten@nokia-bell-labs.com}
\and
\IEEEauthorblockN{Frederik Vandeputte}
\IEEEauthorblockA{
\textit{Nokia Bell Labs}\\
Antwerp, Belgium \\
frederik.vandeputte@nokia-bell-labs.com}
\and
\IEEEauthorblockN{Tom Van Cutsem}
\IEEEauthorblockA{
\textit{Nokia Bell Labs}\\
Antwerp, Belgium \\
tom.van\_cutsem@nokia-bell-labs.com}
}

\maketitle

\begin{abstract}
We consider the problem of developing suitable learning representations (embeddings) for library 
packages that capture semantic similarity among libraries. Such representations
are known to improve the performance of downstream learning tasks (e.g. classification)
or applications such as contextual search and analogical reasoning.

We apply word embedding techniques from natural language processing (NLP) to train 
embeddings for library packages (``library vectors''). Library vectors represent libraries by 
similar context of use as determined by import statements present in source code. Experimental 
results obtained from training such embeddings on three large open source software corpora reveals 
that library vectors capture semantically meaningful relationships among software libraries, such 
as the relationship between frameworks and their plug-ins and libraries commonly used together 
within ecosystems such as big data infrastructure projects (in Java), front-end and back-end web 
development frameworks (in JavaScript) and data science toolkits (in Python).
\end{abstract}


\section{Introduction}

Over the last few decades we have witnessed an exponential growth in the total lines of available
open source code~\cite{deshpande08total}. In recent years, we also witness exponential growth in 
the total number of library packages or modules (hereafter called ``libraries'') 
available for specific programming languages\footnote{The website \url{http://www.modulecounts.com} 
provides an up-to-date view on the size and growth of the most popular package repositories per 
language.}.

While such large open source software ecosystems are generally a boon to developer productivity, they 
introduce a new set of problems. One such problem is that finding the most relevant library for a 
given task in a specific developer context is growing more challenging. One way in which developers 
deal with this discoverability problem is by publishing and sharing curated lists of libraries, as 
witnessed by the popular ``awesome lists'' on GitHub\footnote{See e.g. 
\url{https://github.com/sindresorhus/awesome} for a meta-list of ``awesome lists''.}. While these 
lists certainly are valuable due to their curation by a large developer community, they hardly form 
a scalable way of indexing today's open source ecosystems. As one data point, the website 
\url{www.libhunt.com} which is based on such lists has indexed on the order of 23,000 projects 
whereas according to \url{www.modulecounts.com} the total count of the top 6 most popular package 
repositories well exceeds over 1 million packages (data as of January 2019) and so most packages in 
the ``long tail'' remain undiscoverable through manual curation.

The size and scale of today's software ecosystems suggests that a machine learning approach could 
help us build tools that help developers more effectively navigate them. However, for most learning 
algorithms to be applied successfully to this problem, we require a mathematical representation of 
libraries, preferably one that represents similar libraries by similar representations.

This paper addresses the question whether we can leverage techniques from natural language 
processing, in particular word embeddings, to learn meaningful distributed representations of software 
libraries from large codebases. Just like word embeddings learn to represent similar words by 
similar dense vector representations based on the words' similar context of use, we aim to learn a 
dense vector representation of libraries based on their context of use, 
which in this paper we compute by looking at how they are imported alongside other libraries across source files and projects.

Our work positively answers this research question. We describe how Mikolov et. al's skip-gram 
model~\cite{mikolov13distributed}, which is used to learn embeddings for words (``word vectors'') 
can be adapted to learn embeddings for libraries (``library vectors'') based on their context of 
use. Just like word vectors are trained on large corpora of natural language text, library vectors 
require large corpora of source code. We trained library vectors for three software ecosystems 
(JavaScript, Java and Python\footnote{Today's three top languages as measured by number of 
contributors on GitHub (\url{https://octoverse.github.com/projects\#languages}, accessed January 
2019) as well as by total number of packages in their most popular package managers 
(\url{modulecounts.com}, accessed January 2019).}) for which we report a detailed quantitative and 
qualitative analysis. Finally, we show how the trained embeddings can be used as the basis for a 
contextual search engine where queries can be ``anchored'' on one or more known libraries in order 
to tailor the search results to a specific software development context, with the goal of making 
the results more relevant.

\textbf{Supplementary material} We share datasets of trained library vectors together with a Jupyter notebook to explore the data (cf. Section~\ref{sec:evaluation}) at \url{https://zenodo.org/record/2546488}.
\section{Semantic representations for Libraries}
\label{sec:semantic_representations}

If we were able to cluster libraries by their semantic similarity, we would be able to
more effectively categorize a larger fraction of the growing library ecosystems, in turn helping
developers find more relevant libraries. But how can ``semantic similarity'' be defined for libraries? There are many valid interpretations: the 
libraries could offer the same functionality, could both be extensions of the same base framework, 
could follow the same API guidelines, could be written by the same author, and so on.

Another measure of semantic similarity, and one that we will build on in the rest 
of this paper, is to consider the libraries' context of use: the set of other libraries that are 
also frequently used by code that imports the target library. By this measure of similarity, Python 
libraries such as NumPy and SciPy would be very close, as SciPy (a toolkit for scientific 
programming) builds on the data types provided by NumPy (e.g. N-dimensional arrays).
More generally, we conjecture that libraries would get clustered around platform libraries but also around more abstract concepts such as data processing, machine learning, visualization, networking, etc.

To identify the co-occurrence of libraries in source code one can look specifically at
import statements, which are often used in very idiomatic ways across projects and languages.
Consider the following Python script:

\begin{lstlisting}[language=Python]
import numpy as np
from scipy import linalg
A = np.array([[1,2],[3,4]])
linalg.inv(A) # invert matrix A
\end{lstlisting}

Parsing the file and post-processing the import statements reveals that libraries ``numpy'' and 
``scipy'' co-occur in this source file. If this combination repeats in many source files across 
different projects, it is reasonable to infer that the libraries are closely related. 
Table~\ref{tab:similarity_examples} lists examples from a few domains across different library 
ecosystems.

\begin{table}[ht]
  \caption{Examples of semantic similarities among libraries} \label{tab:similarity_examples}
  \centering
  \resizebox{\columnwidth}{!}{
  \begin{tabular} { l l l } 
    Domain & Example library & Libraries with similar usage context\\
    \hline
    \hline
    \multicolumn{3}{l}{Java (Maven artifacts)} \\
    \hline
    \makecell[l]{Big Data} & \makecell[l]{org.apache.spark}     & \makecell[l]{hive, hbase, hadoop, kafka}\\
    \makecell[l]{Web App} & \makecell[l]{org.springframework}     & \makecell[l]{hibernate, elasticsearch, jackson-core}\\
    \makecell[l]{NLP} & \makecell[l]{org.apache.opennlp}     & \makecell[l]{stanford-corenlp, lucene, uima}\\
    \hline
    \multicolumn{3}{l}{JavaScript (NPM modules)} \\
    \hline
    \makecell[l]{Front-end Dev} & \makecell[l]{jquery}     & \makecell[l]{bootstrap, d3, backbone, leaflet}\\
    \makecell[l]{Front-end Dev} & \makecell[l]{react}     & \makecell[l]{react-dom, react-router, redux, immutable}\\
   \makecell[l]{Web App} & \makecell[l]{express}     & \makecell[l]{request, morgan, socket.io, cors}\\
    \hline
    \multicolumn{3}{l}{Python (PyPI packages)} \\
    \hline
    \makecell[l]{Data science} & \makecell[l]{numpy}     & \makecell[l]{scipy, matplotlib, pandas, sklearn}\\
    \makecell[l]{NLP} & \makecell[l]{nltk}     & \makecell[l]{gensim, joblib, spacy, textblob}\\
    \makecell[l]{Web App} & \makecell[l]{django}     & \makecell[l]{django-extensions, celery, pygments, wagtail}\\
  \end{tabular}}
\end{table}

The idea of considering context to determine similarity is directly based on
the word2vec word embedding model~\cite{mikolov13distributed},
which learns embeddings for words by looking at 
neighbouring words that co-occur with the target word. The conjecture is that similar words will 
tend to occur with similar context words. Word embeddings have been shown to capture semantic and 
syntactic similarities of words~\cite{mikolov13distributed}, for example vec(``Germany'') $+$ 
vec(``capital'') is close to vec(``Berlin'') and vec(``quick'') is close to vec(``quickly''). Word 
vector arithmetic has been used to engage in analogical reasoning of the form ``$a$ is to $a^*$ as 
$b$ is to \_\_''. For example, the analogy ``king is to queen as man is to woman'' can be encoded 
by the equation vec(``king'') $-$ vec(``queen'') $=$ vec(``man'') $-$ vec(``woman'').

Word embeddings have been used to great success in natural language processing to boost the 
accuracy of downstream learning tasks, such as sentence classification~\cite{kim2014convolutional}. 
The premise for our work is that a similar representation for libraries is likely to provide a 
similar boost in performance compared to a standard sparse ``one-hot'' encoding of libraries for 
learning tasks that involve reasoning about libraries. In addition, embeddings for libraries that 
capture semantic similarity enable applications such as contextual search, analogical reasoning and 
automated categorization of libraries.

We go into more detail on how library vectors are trained in Section~\ref{sec:training_model}. We 
first introduce the datasets on which we have trained the library vectors. Just like 
word embeddings rely on co-occurrence patterns in natural language text, library embeddings rely on 
co-occurrence patterns in source code. To see if useful library vectors can be trained, we 
need to understand the patterns related to importing libraries in code.
\begin{table*}[!ht]
\caption{Basic statistics of all crawled open source projects} \label{tab:datasourcestats}
\begin{tabular} { l | c c | c c | c c | c c | c c | c c}
  Language & \multicolumn{2}{c|}{Java} & \multicolumn{2}{c|}{JS} & \multicolumn{2}{c|}{Python} & \multicolumn{2}{c|}{Ruby}  & \multicolumn{2}{c|}{PHP} & \multicolumn{2}{c}{CSharp} \\
  Repo &  MVNCentral & GitHub & NPM & GitHub & PyPi & GitHub  & RubyGems & GitHub  & Packagist & GitHub  & Nuget & GitHub \\ \hline
\# Projects &  120K & 260K & 380K & 310K & 110K & 260K & 130K & 110K & 130K & 69K & 100K & 93K \\
\# Removed Clones &  3\% & 2.4\% & 19\% & 9.2\% & 9.7\% & 5.8\% & 5.5\% & 12\% & 12\% & 8.6\% & 8.7\% & 3.2\% \\
\# Source Files &  5M & 24M & 3.8M & 6.5M & 1.5M & 7.6M & 1.3M & 2.7M & 2.7M & 4.8M & 5M & 8.3M \\
\# Import Stmts &  38M & 200M & 16M & 34M & 9.3M & 51M & 3.1M & 5.8M & 8.6M & 15M & 17M & 37M \\
\# Unique Imports &  2M & 8.6M & 1.4M & 2.5M & 1.6M & 3.6M & 690K & 570K & 1.1M & 870K & 310K & 540K \\\end{tabular}
\end{table*}

\begin{table*}[t]
\caption{Raw imports extracted from 5 representative source files of the pytorch CycleGAN-and-pix2pix GitHub project.} \label{tab:importsexample}
\begin{tabulary} {\linewidth}{lL}
Source file & Imports \\
\hline
data/single\_dataset.py & os.path, data.base\_dataset.BaseDataset, data.base\_dataset.get\_transform, data.image\_folder.make\_dataset, PIL.Image \\
models/cycle\_gan\_model.py & torch, itertools, util.image\_pool.ImagePool, base\_model.BaseModel \\
models/networks.py & torch, torch.nn, torch.nn.init, functools, torch.optim.lr\_scheduler \\
util/get\_data.py & \_\_future\_\_.print\_function, os, tarfile, requests, warnings.warn, zipfile.ZipFile, bs4.BeautifulSoup, os.path.abspath, os.path.isdir, os.path.join, os.path.basename \\
util/visualizer.py & numpy, os, sys, ntpath, time, scipy.misc.imresize \\
\end{tabulary}
\vspace{-2mm}
\end{table*}

\section{Analysis of Library Imports in Big Code}
\label{sec:importanalysis}

We analyzed the imports in open source software projects across 6 languages to better understand their co-occurrence and reuse patterns, answering the following key questions:

\begin{enumerate}
\item How to appropriately clean the raw data?
\item How many relevant imports does each ecosystem have?
\item At what granularity should import vectors be trained?
\item How (often) do imports co-occur with each other?
\end{enumerate}


\subsection{Data Acquisition and Extraction}

We analyzed the imports of 6 languages, as depicted in Table~\ref{tab:datasourcestats}.
For each language, we crawled both all relevant open source projects on GitHub as well as all library packages on their main public package repository.
We only fetched GitHub projects with at least 2 stars to significantly reduce\footnote{Roughly 10\% of the 4M Java/JS GitHub projects have at least 2 stars.~\cite{githubsearch}} the number of projects to crawl and preprocess, while still retaining a statistically representative subset of 100k-400k projects.
As GitHub is known to contain significant amounts of duplicate code~\cite{lopes2017dejavu}, we also removed projects that import exactly the same set of imports across all source files (projects which we dub {\it import clones}). This resulted in 2.4--19\% reduction in projects across languages (see Table~\ref{tab:datasourcestats}).
We will later analyze the impact of applying additional quality filters.

For the remaining projects, we extracted all language-specific source files, resulting in (tens of) millions of source files per package repository containing at least one import statement.
Given the fact that the GitHub projects often also contain application projects (whereas the public repositories mostly contain library packages), it is not surprising that there are on average more source files per GitHub project than per public repository package.

\begin{figure}
  \centering
  \includegraphics[width=0.49\textwidth]{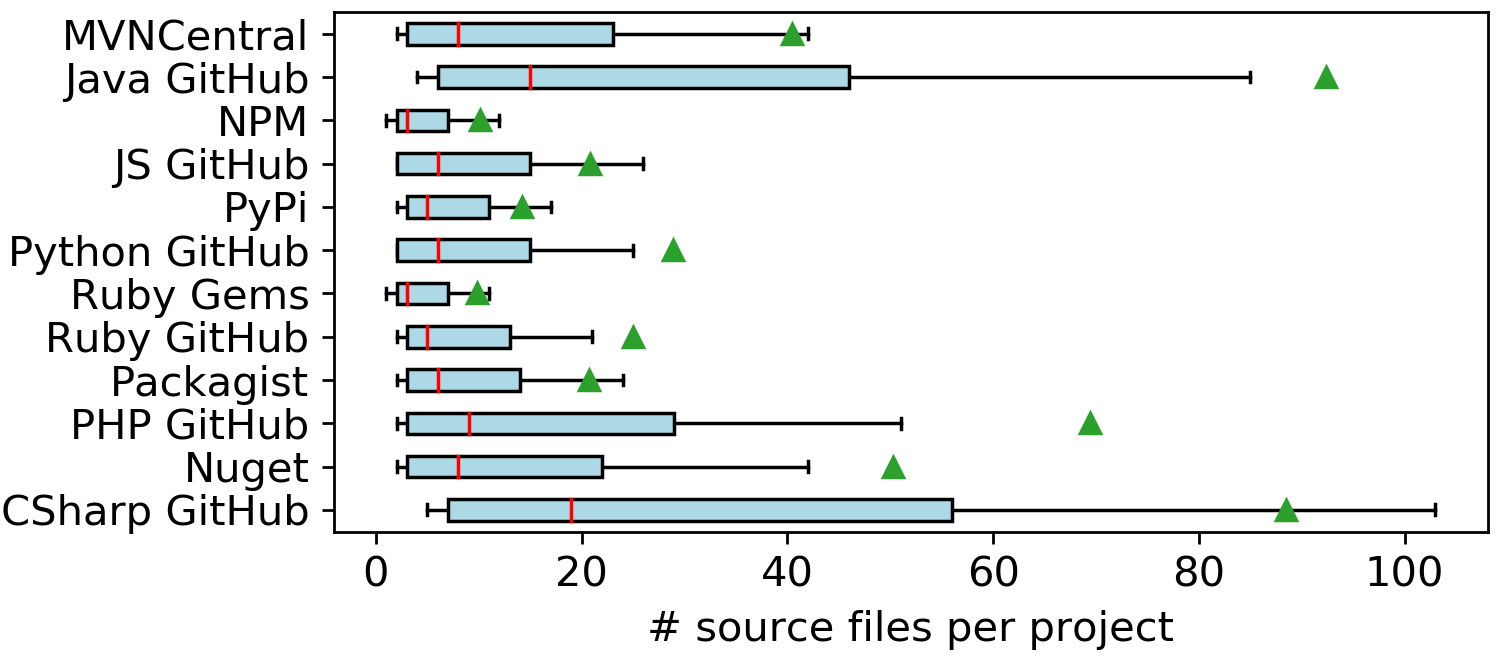}
  \caption{Boxplots of the number of source files per project.}\label{fig:nsourcefilesdistro}
\vspace{-3mm}
\end{figure}

In Figure~\ref{fig:nsourcefilesdistro}, we summarized the distribution of the number of source files across all projects (the green triangle represents the mean).
Unless otherwise stated, the whiskers in each boxplot represent the 15th and 85th percentiles.
As to be expected, these distributions are heavily right-skewed (i.e. the mean is larger than the median), with most projects being very small in size and a few projects being extremely large.
For example, the top 10 largest projects per repository typically contain several (tens of) thousands of source files (up to 140k source files for the largest project,  Glassfish). 
As such, we do not show the outliers on these boxplots for clarity.
Some other observations from this plot, are that GitHub projects are typically larger and have more variation in size than those from the public package repository, and  that NPM and RubyGems libraries, which often contain a single function, are very small compared to those from other repositories like Maven or Nuget, which often contain many classes and packages.


Next, as we are only interested in the import dependencies, we create a derived dataset from these raw source files, extracting only the language-specific import statements. 
The methodology differs slightly for every language and is described in more detail in the Appendix. We only considered statically declared import statements and made no attempt to
capture dynamic library loading or runtime code injection.

This results in a set of raw imports that is retained for every source file for all projects.
In our example Python code snippet, the raw imports extracted would be ``numpy'' and ``scipy.linalg''. We post-process these tokens to map to a canonical library name before learning their vector representation (see Section~\ref{sec:importgranularity}). For a more realistic
example of the kind of raw imports found in our corpus, Table~\ref{tab:importsexample} shows the raw imports extracted from a representative subset of source files from the pytorch CycleGAN GitHub project~\cite{exampleproject}.
Source files often consist of a mixture of external and internal imports (e.g. ``torch.nn'' and ``base\_model.BaseModel'', respectively), co-occurring in different patterns across different source files.

\begin{figure}
  \centering
  \includegraphics[width=0.49\textwidth]{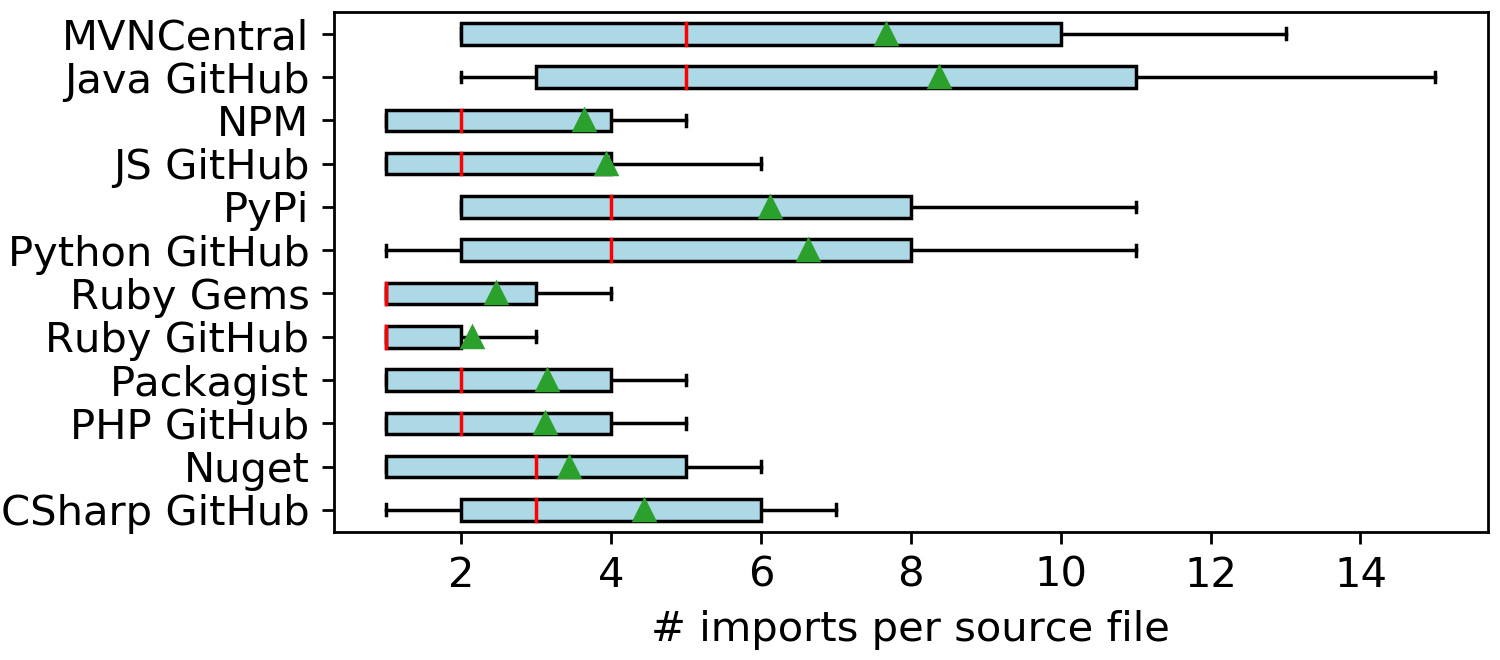}
  \caption{Boxplots of the number of imports per source file.}\label{fig:nimportsdistro}
\vspace{-3mm}
\end{figure}


\begin{figure*}[t]
  \centering
  \begin{subfigure}[b]{0.49\textwidth}
    \includegraphics[width=\textwidth]{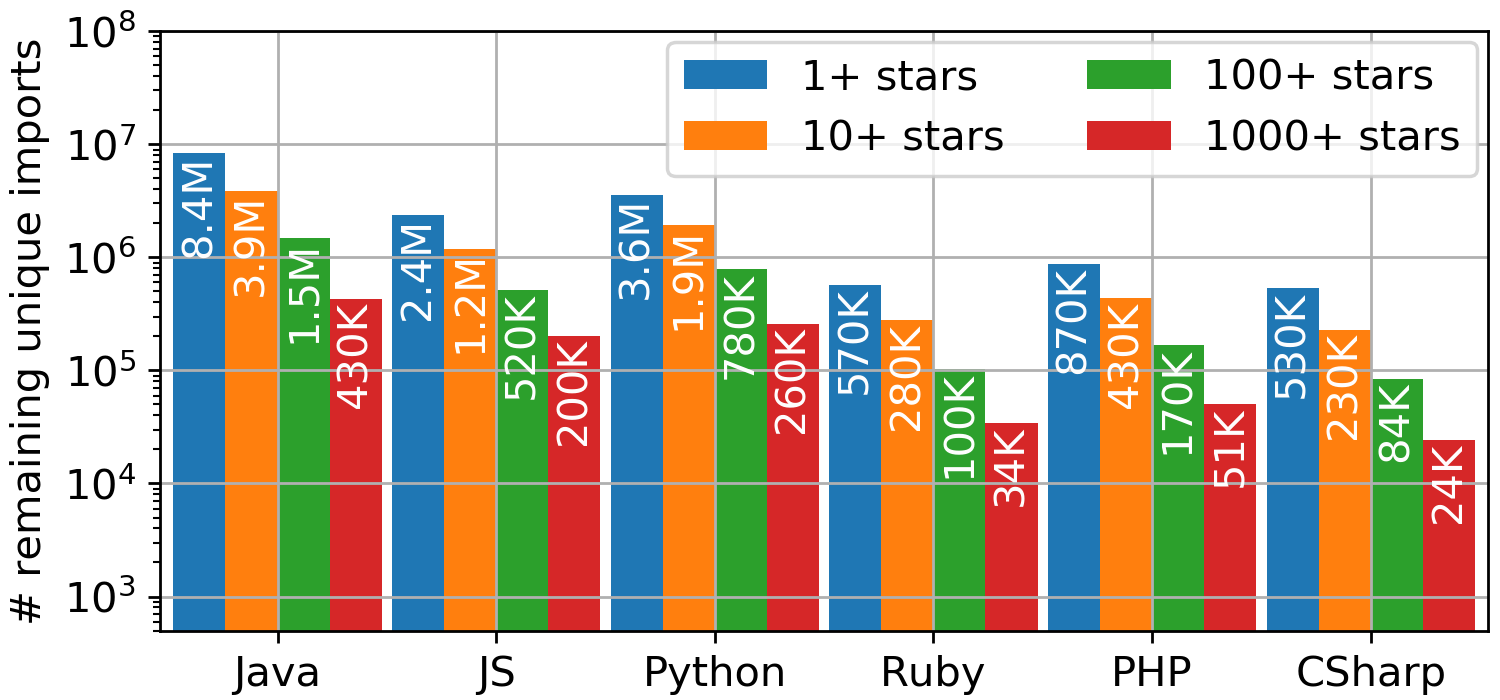}
    \caption{\# imports with global reuse $R>=1$}
    \label{fig:reuse_versus_stars_1}
  \end{subfigure}
  \begin{subfigure}[b]{0.49\textwidth}
    \includegraphics[width=\textwidth]{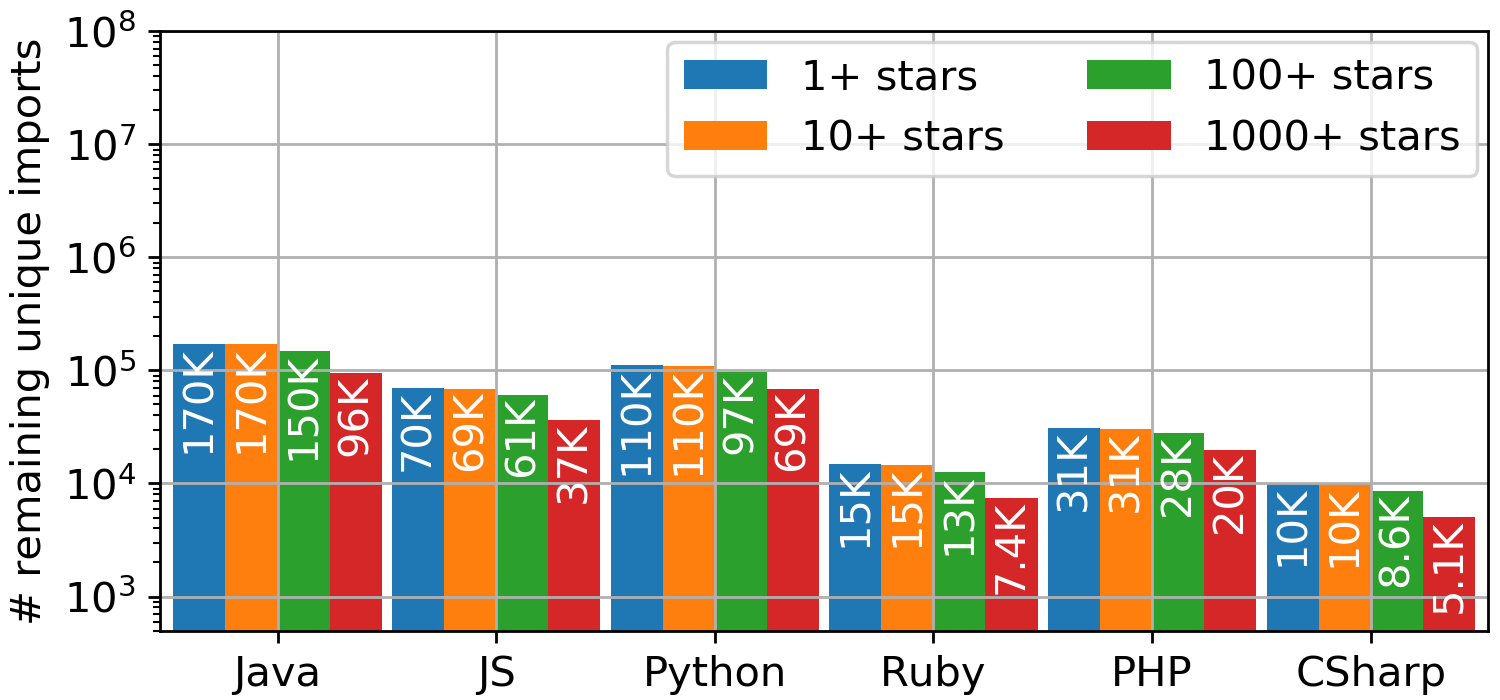}
    \caption{\# imports with global reuse $R>=10$}
    \label{fig:reuse_versus_stars_10}
  \end{subfigure}
  \caption{Impact of number of GitHub stars on \# imports with minimal global import reuse $R$.}\label{fig:reuse_versus_stars}
  \vspace{-3mm}
\end{figure*}

The total number of import statements extracted from this dataset is also shown in Table~\ref{tab:datasourcestats}.
As to be expected, the total number of import statements is a multiple of the number of source files.
Figure~\ref{fig:nimportsdistro} shows the distribution of import statements per source file across all projects.
Interestingly, the distributions are very similar within the same language, even though GitHub projects are on average much larger than their public repository libraries.
The differences between languages can be explained due to different import granularities (e.g. class-level Java imports such as \texttt{import java.util.List} versus library-level NPM imports such as \texttt{require('fs')}) and different programming styles and code reuse patterns.



Finally, from Table~\ref{tab:datasourcestats}, we can also observe that the number of unique imports is only a fraction of the total import statements, as imports are typically repeatedly used in multiple source files of the same project (i.e., {\it local reuse}), as well as across multiple projects (i.e., {\it global reuse}). 
We will further analyze this in the next section.







\subsection{Import filtering: Extracting relevant imports}
\label{sec:importfiltering}

In this work, we focus on building vector representations of library imports that are reused across multiple projects (i.e. global reuse). We do not concern ourselves here with
imports that are only reused inside a single project.
A fundamental question is to find a minimal reasonable global reuse threshold for filtering {\it relevant} imports.
Ideally, we want to keep both the popular as well as the specialist high-quality niche imports.

\begin{figure}
  \centering
  \includegraphics[width=0.49\textwidth]{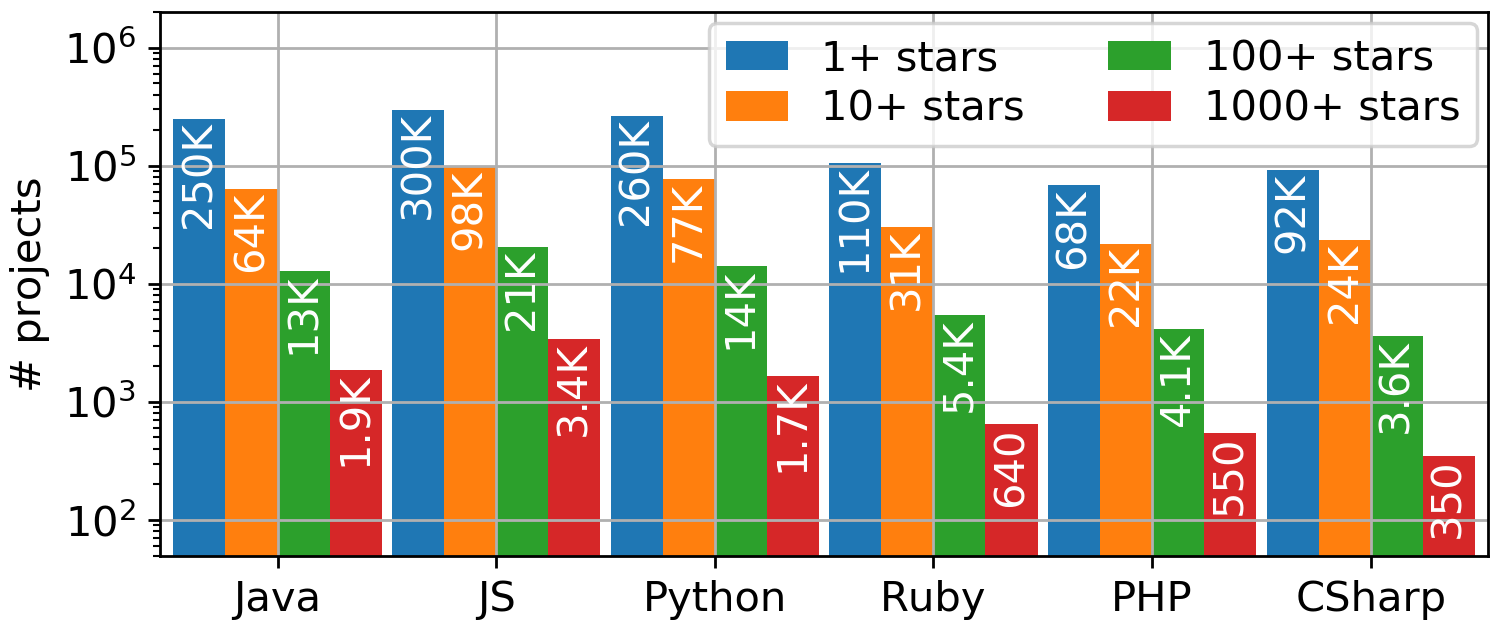}
  \caption{Remaining GitHub projects with at least {\it S} GitHub stars.}\label{fig:nprojects_versus_stars}
  \vspace{-3mm}
\end{figure}

\begin{figure*}[t]
  \centering
  \includegraphics[width=1\textwidth]{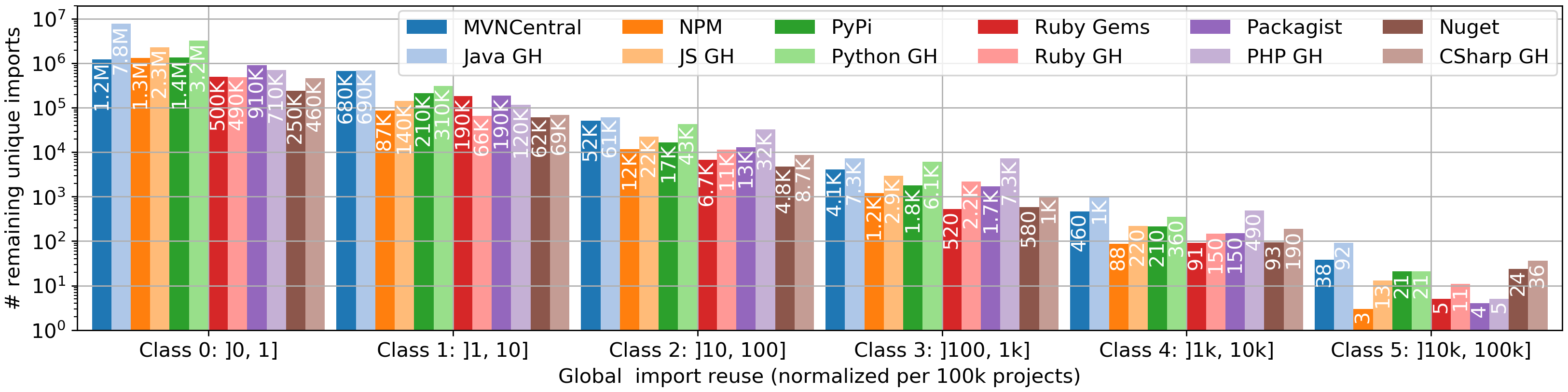}
  \caption{Number of unique imports per import popularity class.}\label{fig:reuse_versus_freq}
  \vspace{-3mm}
\end{figure*}

We investigated how many of the imports also appear in projects with a particular number of GitHub stars.
Figure~\ref{fig:reuse_versus_stars} shows the impact of the number of GitHub stars on the number of imports that are reused across at least {\it R} projects and are also used in at least one project with at least $S$ stars.
Figure~\ref{fig:nprojects_versus_stars} depicts how the number of projects with at least $S$ GitHub stars
drastically decreases by a factor $3-6\times$ for every $10\times$ increase of $S$.
When all imports are considered (i.e. $R>=1$), the number of remaining imports drops by a factor $2-3\times$ for every $10\times$ increase of $S$. 
In contrast, when only considering imports with a global reuse of $R>= 10$, the number of remaining imports drops by $0-15\%$, and by up to $50\%$ for $S>=1000$, which is still remarkable given the tiny fraction of remaining projects with at least 1000 stars.

As such, as a rule-of-thumb, we will only consider imports with a global project reuse $R>=10$.
Empirically, we verified that this threshold works well aross all 6 datasets, even though they significantly vary in size.
Note that we don't want to just retain the imports of highly-starred projects, as imports in high-starred projects are not the only relevant imports.
Though our simple heuristic inevitably will retain some low-quality imports, it is generic and can be used also for the datasets for which we have no associated GitHub stars or reliable ratings.

Next, we analyze the popularity distribution of imports.
A {\it popular} import is an import with a high global reuse across projects.
Figure~\ref{fig:reuse_versus_freq} depicts the number of unique imports per {\it import popularity class}.
We define an import popularity class as all imports with a global reuse of a particular order of magnitude.
Note that we normalized these classes per 100k projects to be able to properly compare the different languages and repositories.

Several key observations can be made from this.
First, one can observe a negative linear trend across all languages on this log-log plot, meaning that there are roughly 10 times fewer imports that are 10 times more popular, or vice versa.
In fact, the popularity-count distribution can be represented as a simple monomial $y = a.x^k$ with exponent $k\approx -1$ and scale factor $a\approx10^7$.
Secondly, the number of imports per popularity class is similar in terms of orders of magnitudes across all languages, though there are also some clear deviations. 
For example, the most popular import class (i.e., class 5) deviates from this pattern, with notable differences between languages.
These differences are foremost caused by the fact that Java, Python and CSharp have an extensive set of runtime libraries with many class-level imports, whereas JavaScript for example only has a narrow set of core libraries.
There are also some differences between the languages regarding the relative amount of imports of medium popularity.

\begin{figure}
  \centering
  \includegraphics[width=0.49\textwidth]{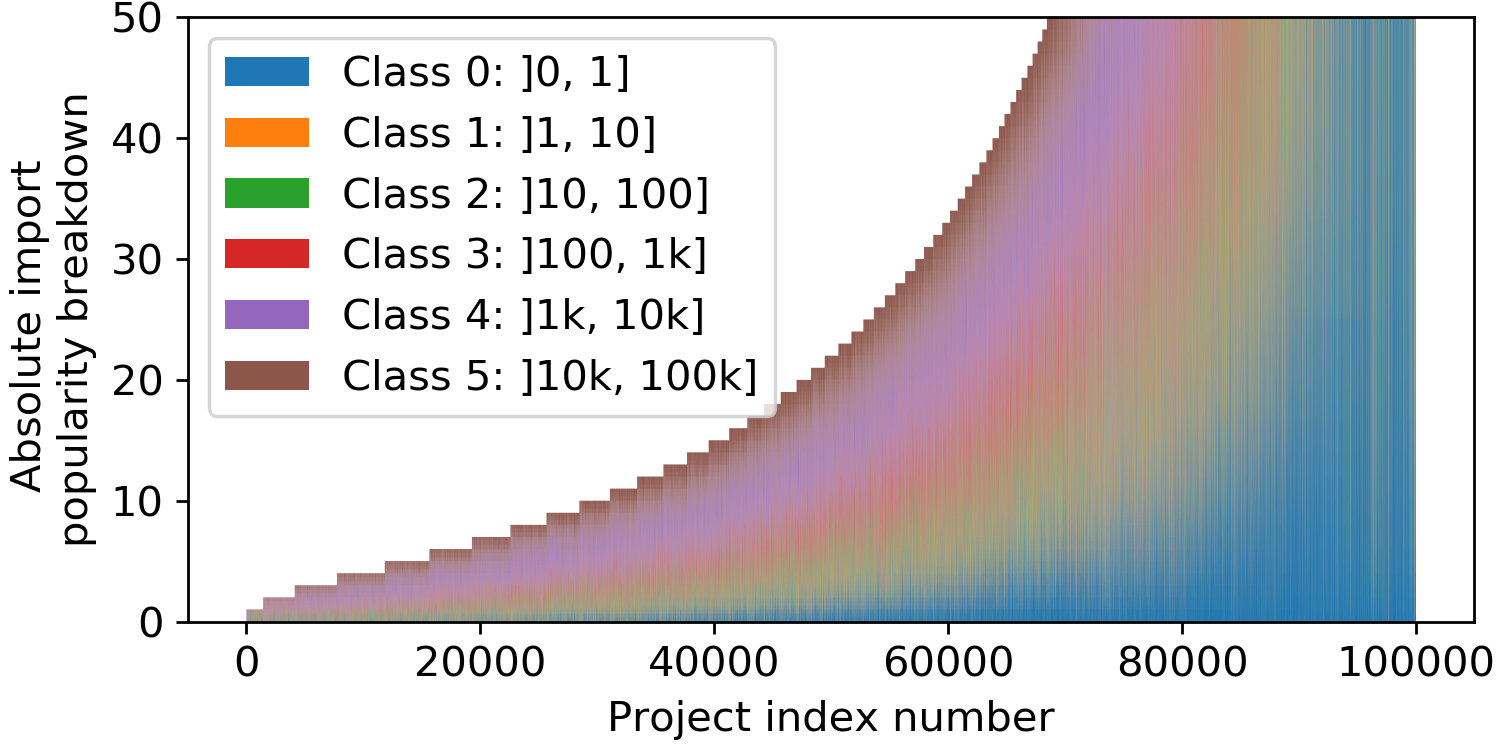}
  \caption{Absolute import popularity breakdown per project for the top 100k JavaScript projects with the most GH stars.}
  \label{fig:popularitybreakdown}
\end{figure}

Figure~\ref{fig:popularitybreakdown} shows for each of the top 100k JS GitHub projects, the breakdown of all unique imports by their popularity class.
In this graph, the projects where partially ordered based on their import popularity class to better highlight the import popularity distribution variations across projects across different sizes.
Several observations can be made:
\begin{itemize}
\item Most projects have imports of all popularity classes;
\item Projects significantly vary in size, with the 15\% smallest projects mostly importing very popular libraries , and the 10\% largest projects mostly importing local libraries.
\item The imports of classes 2--4, which are neither the superpopular nor the project-specific imports, represent a significant fraction of all imports.
\end{itemize}

\begin{figure}
  \centering
  \includegraphics[width=0.49\textwidth]{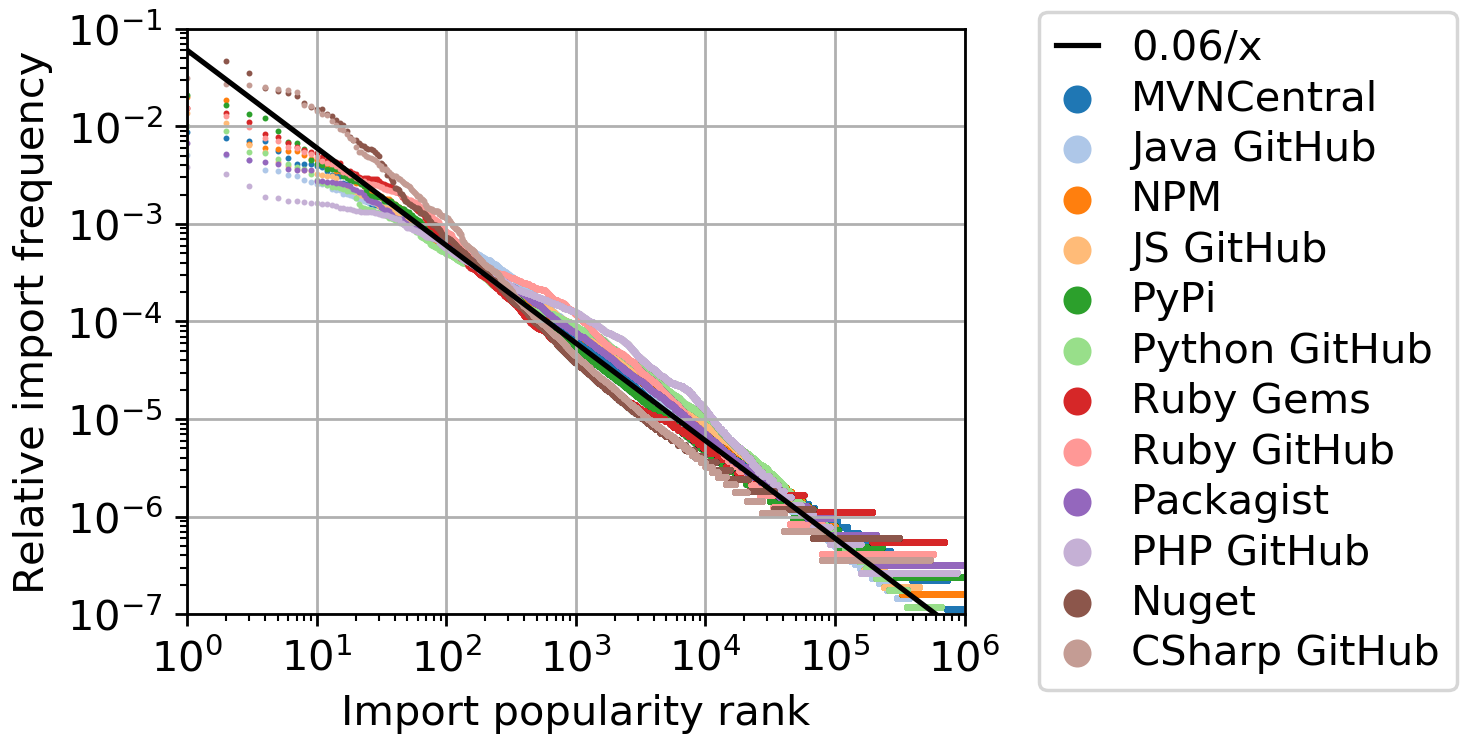}
  \caption{Import popularity distribution roughly obeys Zipf's law.}\label{fig:zipf}
\end{figure}

In Figure~\ref{fig:zipf} we plot the import popularity rank with respect to their relative frequency.
The imports of all 6 programming languages clearly obey Zipf's law~\cite{zipf29freq} (cfr. the reciprocal function $0.06/x$), which is often also used to describe the frequency of words in natural language~\cite{piantadosi14zipf}, web traffic~\cite{adamic2011zipf}, etc.
Some of the deviations in the top and middle sections have been addressed earlier (see the analysis of Figure~\ref{fig:reuse_versus_freq}).
 


We stated earlier that $R>=10$ is a good threshold for filtering out imports that are not 
sufficiently widely used. At the other end of the spectrum, class 5 imports (which appear in more 
than 10\% of all projects and typically represent the runtime libraries of each language)
are generally not very relevant for applications such as search (see 
Section~\ref{par:contextualSearch}) given that they are generally known by most developers using
the language.
As such, we define {\it relevant} imports to be those with a popularity class between 2 and 4, with class 4 representing  popular imports and class 2 representing {\it surprising} imports. 
From Figure~\ref{fig:popularitybreakdown}, we already know that these represent a significant fraction of all imports of most projects.
Specifically, 66\% of all projects contain at least 50\% relevant imports, with over 80\% of projects containing at least 33\% relevant imports. 
As a result, doing a contextual search or creating a recommendation engine for these imports will be useful for most projects.
Note that because of the Zipf distribution, the bulk of all relevant imports are actually surprising imports (i.e., most imports are in the long tail of imports, see also Figure~\ref{fig:reuse_versus_freq}).


\subsection{Import granularity}
\label{sec:importgranularity}

Depending on the programming language, choices can be made for selecting the right level of granularity for training import vectors. For brevity, we will only focus on Java, JavaScript and Python.
In Java and Python, raw source file imports typically represent individual classes or packages. 
In JavaScript on the other hand, the raw imports often represent entire NPM modules.
Imports at the lowest level of granularity carry the most information about the functionality offered by that import, since a class is much more specific than the library in which it is defined.
However, considering class-level imports will often lead to much larger and more trivial contexts to train vectors on, as different classes from the same package or library are often frequently used together.
Considering imports at package-level or library-level granularity instead will result in increasingly more coarser-grain patterns of functionality, but give insight in which submodules or libraries are often combined to implement some higher-level functionality.

\begin{figure}
  \centering
  \includegraphics[width=0.49\textwidth]{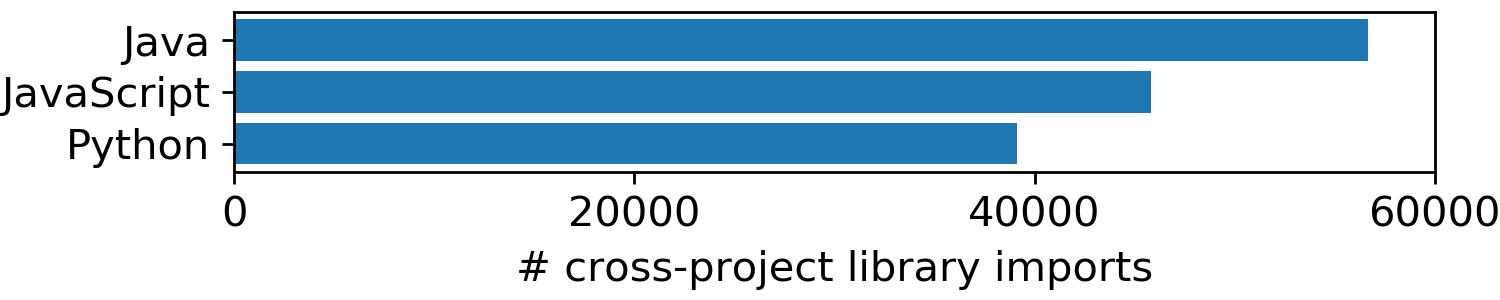}
  \caption{\# unique library imports used for training embeddings.}\label{fig:ntrainedlibimports}
  \vspace{-2mm}
\end{figure}

Selecting the right level of granularity depends on the downstream application.
As we are mainly interested in finding other libraries used in similar high-level contexts, we focus on building models based on library-level imports. 
As such, we first convert our raw import dataset into a corresponding library import dataset, by mapping each import onto the originating library package.
The resulting number of unique library imports we used for training import vectors, is depicted in Figure~\ref{fig:ntrainedlibimports}.
Next, we analyze the co-occurrences of these library imports in our datasets.



\subsection{Library import co-occurrence analysis}

\begin{figure}
  \centering
  \includegraphics[width=0.49\textwidth]{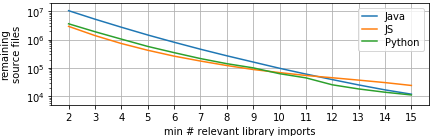}
  \caption{\# source files with at least $R$ relevant library imports.}\label{fig:sourcefiles_with_minimports}
  \vspace{-3mm}
\end{figure}

As we need a sufficient number of import examples to create high quality import vectors, we first analyze the import co-occurrences in source files and projects.
Figure~\ref{fig:sourcefiles_with_minimports} depicts how many source files per language surpass
a given threshold of unique relevant imports, which we define as the {\it context size}.
Roughly 1M source files contain at least 5 unique library imports, and roughly 100k source files contain at least 10 unique library imports.
To be able to train import vectors, we require at least 2 unique library imports per source file. This results in a dataset containing over 30M library imports for Java and over 10M for JS and Python. 


\begin{figure}
  \centering
  \begin{subfigure}[b]{0.49\textwidth}
    \includegraphics[width=\textwidth]{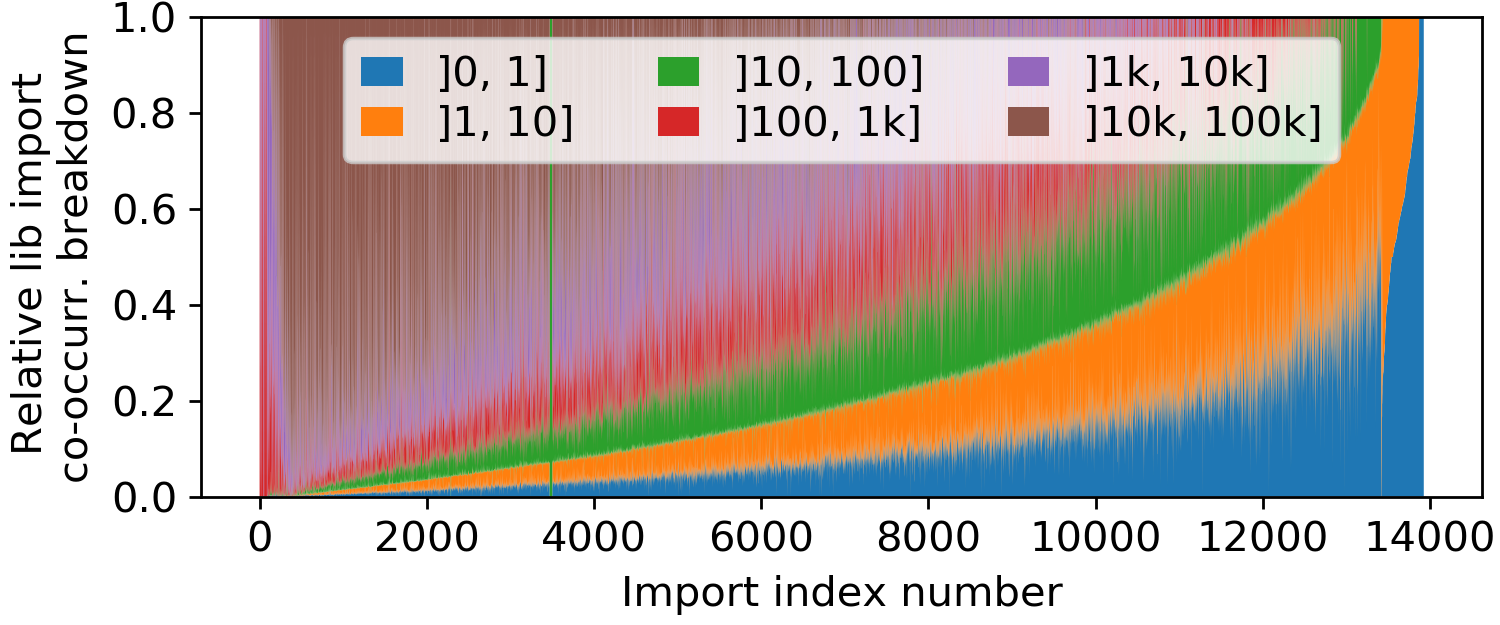}
    \caption{source file co-occurrence frequencies}
    \label{fig:srcfileimportcooccurrencebreakdown}
  \end{subfigure}
  \begin{subfigure}[b]{0.49\textwidth}
    \includegraphics[width=\textwidth]{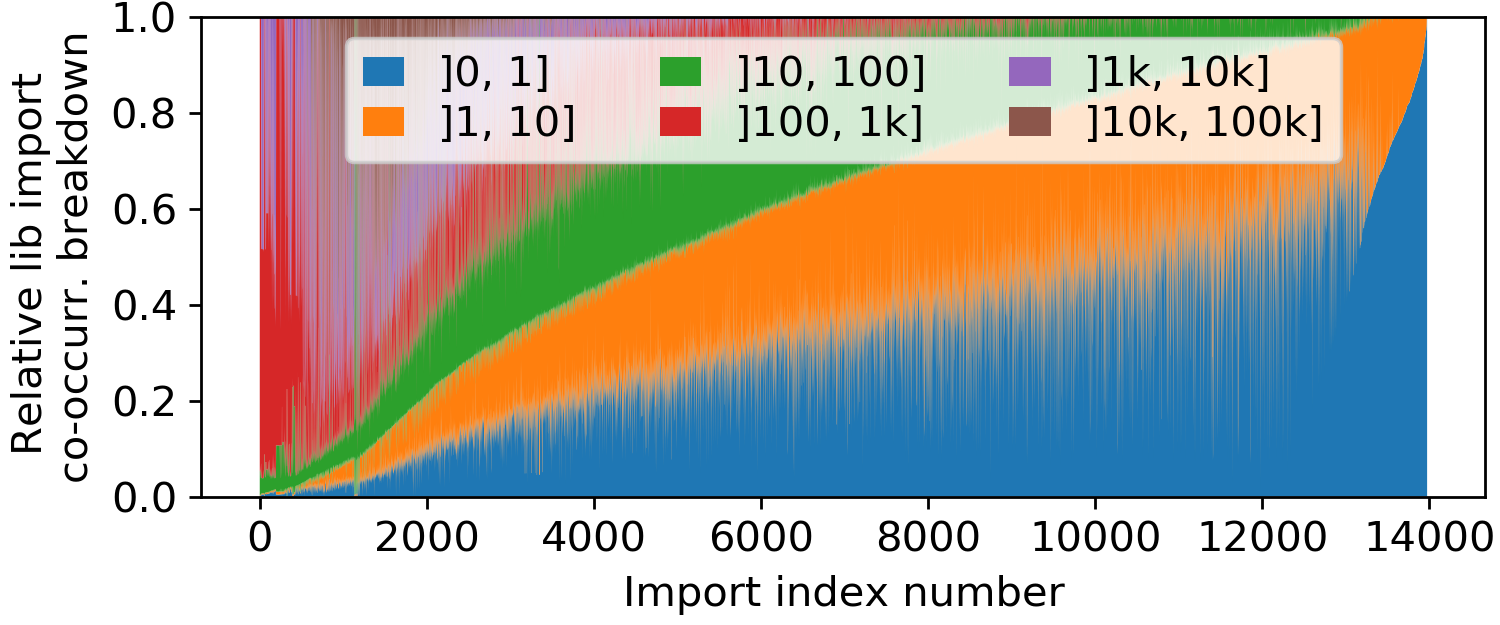}
    \caption{project-level co-occurrence frequencies}
    \label{fig:projectimportcooccurrencebreakdown}
  \end{subfigure}
  \caption{Relative breakdown of co-occurrence frequencies of library imports for the Java dataset.} 
  \vspace{-3mm}
\end{figure}

A second important analysis is whether some import pairs co-occur several times across different source files and projects.
If not, learning import vectors using the skip-gram model would not result in meaningful results.
In our earlier Python example, the ``numpy'' and ``scipy'' libraries are frequently used together, as they are often used in a similar context.
In Figure~\ref{fig:srcfileimportcooccurrencebreakdown}, we show (for the Java GitHub dataset) the relative breakdown of how often each relevant library import co-occurs with other relevant library imports within the same source file.
Here we also partially ordered all library imports by co-occurrence class to better highlight the co-occurrence distribution variations.
As mentioned earlier, imports should ideally co-occur at least several times with other imports.
This is clearly reflected in the results: for 80\% of these library imports, 80\% of all import co-occurrences appear at least twice and 50\% appear at least 10 times.
Note that we observed similar results for JavaScript and Python.

On the other hand, in Figure~\ref{fig:projectimportcooccurrencebreakdown}, we depict the co-occurrence breakdown when considering all import co-occurrences at project level rather than at source file level, meaning all import statements of all source files of a project have been merged into a single set of imports for that project.
As to be expected, the fraction of one-time accidental co-occurrences is much higher, meaning that the cohesion of imports is weaker at project level rather than source file level.
This matches our intuition that imports that co-occur within a source file are typically functionally stronger linked to each other compared to two random project-level imports.
In our earlier example of Python imports (see Table~\ref{tab:importsexample}), modules ``numpy'' and ``requests'' will co-occur less frequently than ``numpy'' and ``scipy''.




\section{Learning Embeddings for Library Imports}
\label{sec:training_model}


Recall from Section~\ref{sec:semantic_representations} that our overall goal is to be able
to cluster libraries based on their semantic similarity, as defined by
their context of use. We indicated that this is equivalent to considering word similarity
based on context of use (i.e. two words are similar if one could be substituted for the other in 
context). This equivalence, combined with the observations from the previous Section regarding 
significant global reuse and library co-occurrence patterns in large corpora of open source code,
suggests that techniques to train a vector representation for words may also be suitable
to train a vector representation for libraries.

We trained embeddings for the three largest ecosystems (JavaScript, Java and Python) based
on the datasets described in Section~\ref{sec:importanalysis}.
To train the embeddings, we consider only imports that are reused in at least two projects.
For downstream tasks such as contextual search, we retain only the vectors of all {\it relevant} imports (cfr. Section~\ref{sec:importfiltering}).

Our dataset consists of triples of the form \textit{(project, filename, [set of imports])}. 
From these triples we generate a [\textit{target-library}, \textit{context-library}] pair for
each combination of 2 libraries imported in the same source file. Thus we only consider
two libraries to co-occur if they are imported within a single source file, not just within
a single project.


\subsection{The Skip-Gram Model} \label{sec:modeltraining}

\begin{figure}
  \centering
  \includegraphics[width=0.45\textwidth]{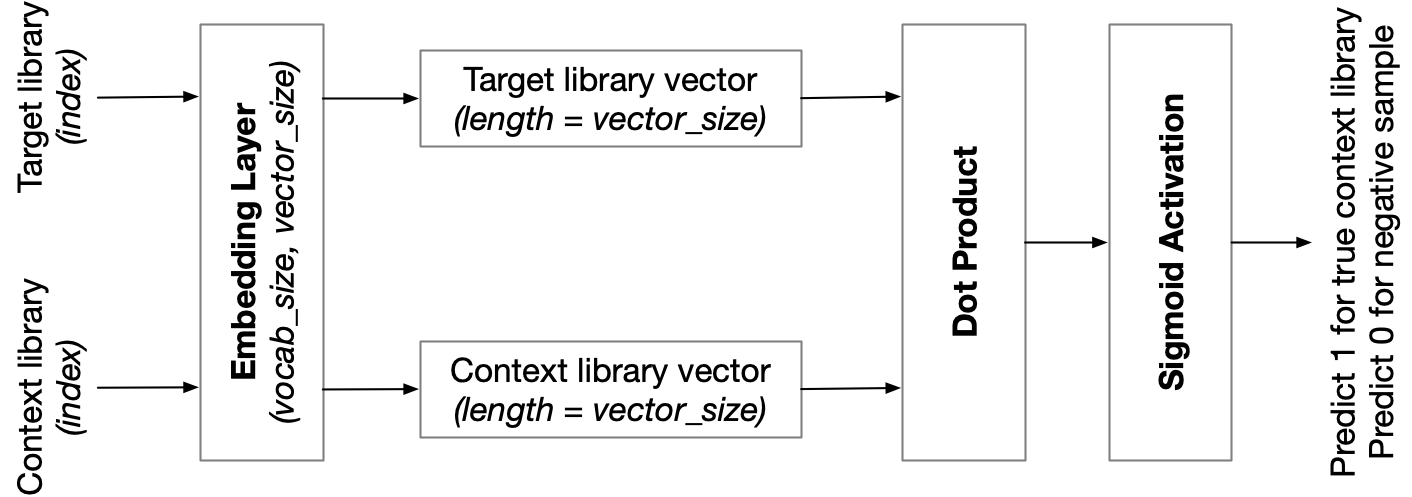}
  \caption{The Skip-gram Model to Train Import Vectors}\label{fig:model}
  \vspace{-3mm}
\end{figure}

To train embeddings we use an architecture based on the skip-gram model with negative sampling 
introduced by Mikolov et. al~\cite{mikolov13distributed} (see Fig. \ref{fig:model}). It is a 
shallow neural network, starting with an embedding layer of size (\textit{number of 
distinct libraries} $\times$ \textit{vector dimensions}), followed by a dot-product and 
a sigmoid activation function.

The goal of the model is to predict, given two library imports,
whether the pair is a positive example, meaning that the \textit{target-library} co-occurred with the \textit{context-library} within a source file of at least one 
project, or a negative example meaning that the libraries were never imported together in any 
source file of any project in the entire dataset.

The positive examples correspond to our preprocessed training data set. To generate negative examples, we randomly sample pairs of libraries from the vocabulary of possible library imports and then verify that these imports never co-occur.
To train the network, we feed it an equal number of positive and negative library pairs.

The training objective of the network is to output 1 when it is fed a positive sample
and to output 0 when it is fed a negative sample. We use binary cross-entropy as the loss
function. As the activation function is preceded by a dot-product of vector embeddings,
to minimize the training loss the network is forced to let vectors for positive library pairs
point in the same direction, and conversely to let vectors for
negative library pairs point in opposite directions. At the end of training, back-propagation
has turned the initially random embedding matrix into semantically meaningful vectors, one vector 
for each library in our vocabulary.

While our skip-gram model is inspired by the model of Mikolov et al. there are
notable differences. The most important difference is that 
Mikolov et al. define the local context of a word using a sliding window over a sentence.
Since the order in which import statements appear in a source file is often insignificant,
it does not make sense to use a sliding window within the bag of libraries imported by a source 
file. Instead we define the entire source file as local context and generate context-target
pairs for each combination of two library imports co-occurring in the same source file.

As it is straightforward to check whether two libraries ever co-occur in any source file,
we make sure that our negative samples are true negatives (i.e. guaranteed to never co-occur).
In Mikolov et al.'s model, negative samples may not be true negatives. We also generate
an equal amount of positive and negative example pairs, as opposed to Mikolov et al. who
report a large negative sampling ratio (from 5 to 20 times more negative than positive examples).

\subsection{Assessing the Quality of the Trained Models}

\begin{figure*}[t]
    \centering
    \includegraphics[width=1\textwidth]{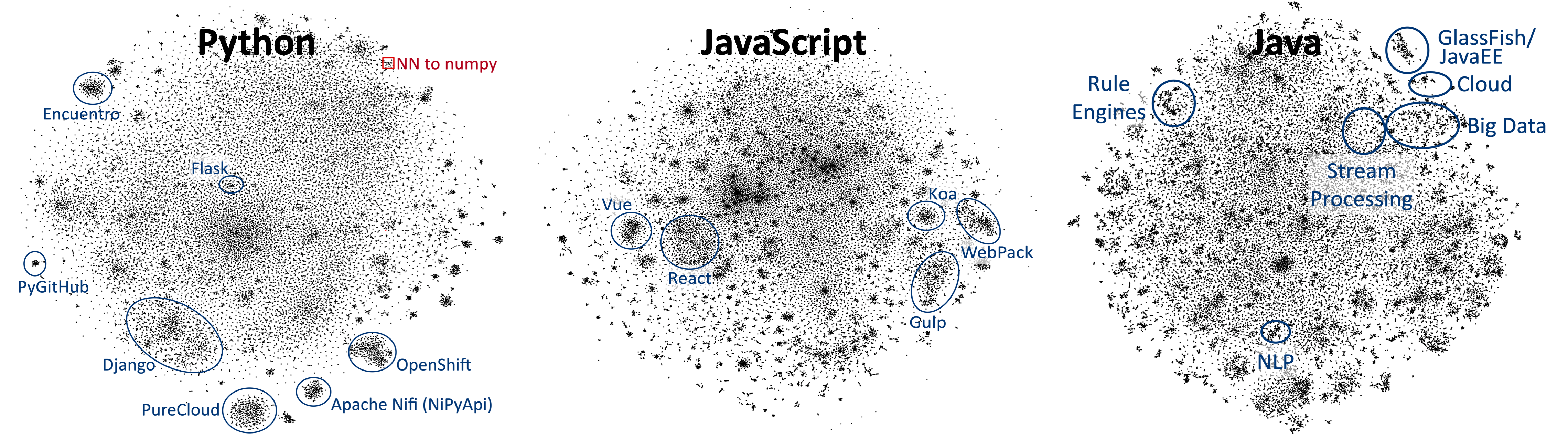}
    \caption{2D-visualization of the Model output for the Python, JavaScript and Java ecosystems}\label{fig:models}
\end{figure*}

As we are dealing with an unsupervised learning task without a reference benchmark it is particularly hard to determine whether a model is of good enough quality and more specifically whether tuning the hyper-parameters of the model make for better results. 
One can visually inspect the embeddings after projecting them down to 2 or 3 dimensions using techniques such as t-SNE~\cite{maaten2008visualizing}. Figure~\ref{fig:models} depicts the 2D t-SNE projection of the three ecosystems under study using a perplexity value of 10 and 3000 iterations.
There are on the order of 40-60K dots in each figure, each dot representing a library. t-SNE identifies clearly isolated clusters, which we manually determined to be large self-contained ecosystems having strong library co-occurrence patterns, such as Apache Nifi, PureCloud or OpenShift. Even in the apparently less clustered areas, local neighbourhoods carry a lot of semantic similarity, as we will discuss later (Fig. \ref{fig:python-numpy}).

Even with such a visual representation interpreting the quality of the embeddings remains a 
subjective and time-consuming task. Library embeddings are harder to interpret than word 
embeddings, as the names of libraries do not always directly convey meaning to a human evaluator. A 
deep understanding of the libraries is often needed to judge their semantic similarity. In 
Section~\ref{sec:evaluation-measure} we will experimentally quantify the quality 
of the learned embeddings.

We trained models at all three levels of import granularity (i.e., class, package and 
library-level) and could clearly observe that, for the task of clustering libraries based on 
semantic similarity, package- and library-level imports produced better non-trivial 
clusters\footnote{Training with class-level imports generates very tight clusters, but they are 
trivial, i.e. the clusters often consist of classes of the same package, and so do not capture 
global reuse patterns.}. In this work we therefore focus on library-level imports.
We experimented with vector dimensions ranging from 10 to 600 and
empirically established that 100 dimensions were sufficient to produce quality vectors. All subsequent results are for 100 dimensions.

\subsection{From Embeddings to Contextual Search} \label{par:contextualSearch}

As the embeddings trained by the skip-gram model are optimized to cluster libraries by
similar context of use, we can exploit this property to perform contextual search: given a
number of libraries (for instance, libraries a developer is already using), find
other libraries relevant in this context.

We built a contextual search engine that takes one or more libraries as input (the developer's 
context), adds up the vectors for those libraries into a \emph{context vector}, and then returns 
the nearest neighbouring libraries to that context vector as the result\footnote{
In practice, the nearest neighbors can be filtered based on additional metadata like tags or keywords. For example, given the library ``org.apache.hadoop'' 
and the tag ``database'', first the nearest neighbours to hadoop can be located and then filtered
for libraries tagged with ``database''. Anchoring on context enables libraries related
to databases such as HBase and Cassandra (often used with hadoop) to be ranked before
databases such as MySQL and PostgreSQL.}.
For this to work we rely on the compositionality of distributed 
representations~\cite{mikolov13distributed}, meaning that the composition of two embeddings
retains semantic properties. Some examples of composing (adding) library vectors are shown in 
Table~\ref{tab:compose}. Examples of nearest neighbours will be shown in the next Section.

\begin{table}[ht]
  \caption{Compositionality of library vectors} \label{tab:compose}
  \centering
  \begin{tabular} { l | l | l }
    \hline
    ecosystem   & context & most notable predictions  \\ 
    \hline
    Java        & postgresql   &  h2database, hsqldb \\ 
                & postgresql + spark & Parquet, Hbase \\ 
    JavaScript  & winston (logging) & log4js \\ 
                & winston + express & morgan (http logging) \\
    Python      & aws + fs & s3 \\ 
  \end{tabular}
\end{table}






\section{Evaluation}
\label{sec:evaluation}

To empirically assess the quality of trained embeddings we show the nearest neighbours of a
number of well-known software libraries, followed by some performance metrics on a modified contextual search algorithm specifically designed for evaluation purposes.
Finally we show how library embeddings can be used in analogical reasoning tasks.


\subsection{Empirical Evaluation}

To explore the learned embedding space we project the embedding space into
two or three dimensions using t-SNE, which is a dimensionality reduction technique that aims to preserve local distances among points~\cite{maaten2008visualizing}.

\begin{figure}[ht]
  \centering
  \includegraphics[width=0.50\textwidth]{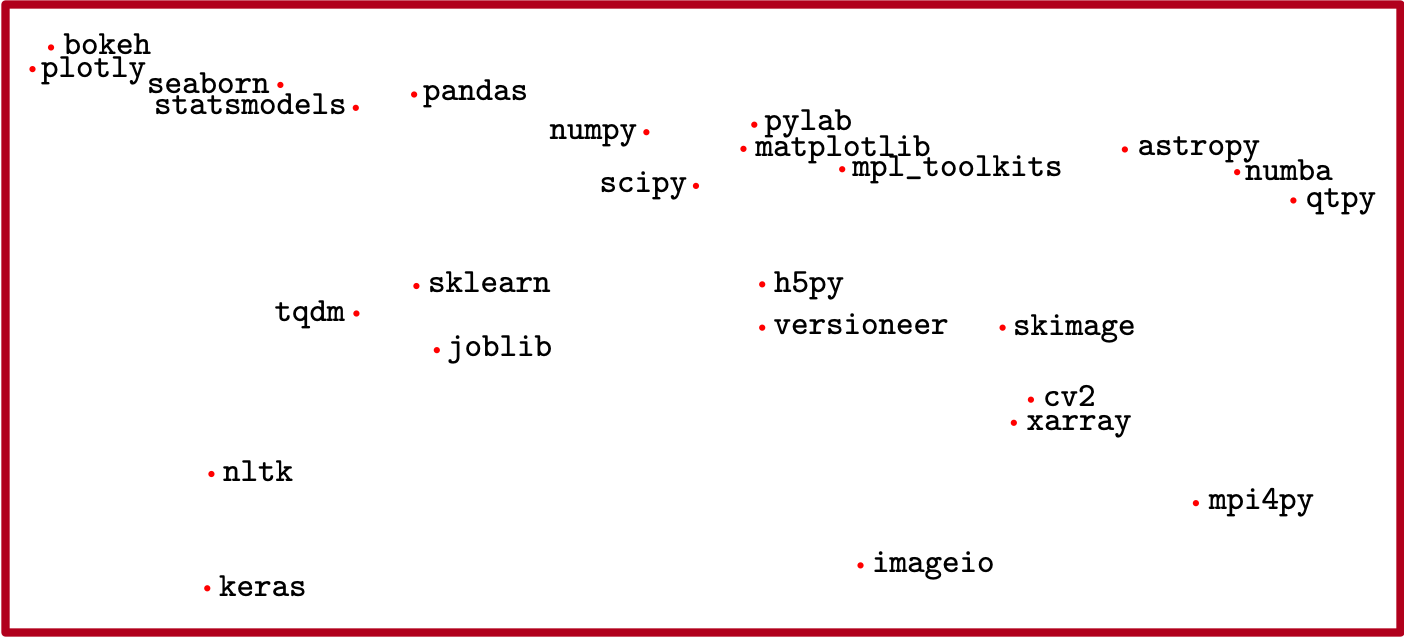}
    \caption{A zoom on the local neighborhood of numpy (Python)} \label{fig:python-numpy}
\end{figure}

As our embedding space consists of tens of thousands of points (one for each library), we additionally developed an interactive visualization tool to pan and zoom the 2D or 3D projection to be able to inspect local neighborhoods. For example, Figure~\ref{fig:python-numpy} shows an unfiltered zoom of the red box in the Python model shown in Figure~\ref{fig:models}. It shows the local neighborhood around the $numpy$ library. The very closest neighbours are $scipy$, $matplotlib$ and $pylab$ which are frequently used together by data scientists. Further out, the neighbours are still relevant, since most of them remain situated in the data science domain. 

\newcommand{\kw}[1]{\tiny \texttt{#1}\footnotesize}

While visual inspection gives us a sense of the local neighbourhood around
specific projects, projecting down from 100 dimensions to 2 or 3 using t-SNE loses information.
To get a more accurate view on the local neighbourhood,
we can calculate the nearest neighbours (NN) to a library vector directly in
the high-dimensional space, using cosine similarity as the distance metric.
Table~\ref{tab:empiricalEval} shows the top-5 nearest neighbours (out of a total of 40-60K libraries per ecosystem) of a number of selected libraries across the three ecosystems under study. For each library we also list its global import popularity rank.
Although we can observe some noise in the results, most of the nearest neighbours are either offering similar functionality, or are often used together. 
In addition to the top-5 nearest neighbour libraries, the table also shows the top-5 most frequently co-occurring libraries to the given library, which we use as a statistical baseline for comparison\footnote{The top-5 most popular standard libraries were filtered out of the baseline results since these would otherwise dominate the results, hiding more relevant ones. These are, for Java: \kw{java.util}, \kw{java.io}, \kw{java.net}, \kw{java.lang.reflect}, \kw{java.util.concurrent}; JS: \kw{path}, \kw{fs}, \kw{http}, \kw{child\_process}, \kw{util}; Python: \kw{setuptools}, \kw{os}, \kw{sys}, \kw{re}, \kw{json}). For Java we only retain a single library with the same Maven groupid.}.

\begin{figure}
  \centering
  \includegraphics[width=0.45\textwidth]{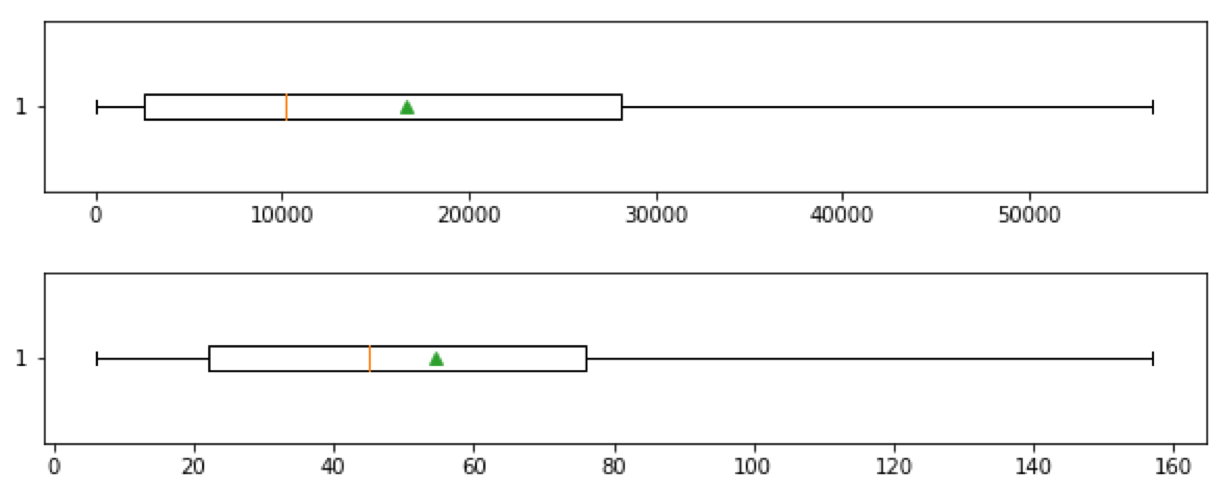}
  \caption{Import Popularity Rank Distribution of NN-search (top) vs. Most Popular Combos (bottom). Note the scales.} \label{fig:pop-dist}

\end{figure}

The examples seem to indicate that library vectors generally surface more surprising (less popular) results compared to the baseline. A more extensive comparison of the popularity rank distribution of top-5 results on a uniform random sample (10\%) of the Java dataset confirms this effect (see Figure~\ref{fig:pop-dist}).

\begin{table*}[ht]
    \caption{Top-5 Nearest Neighbours vs top-5 most frequently co-occurring libraries for some well-known given libraries}  \label{tab:empiricalEval}
    \centering
    \resizebox{\linewidth}{!}{
    \begin{tabular} { |p{0.1cm}| p{2.7cm} p{11cm} p{9cm}| }
    \hline
    & \textbf{Given Library} & \textbf{Top-5 Nearest Neighbours} & \textbf{Top-5 Most Popular combos} \\ 
    \hline
    \hline
    \multirow{7}{*}{\rotatebox{90}{\textbf{Java}}} & \makecell[l]{org.apache.spark \\ \\ } & \makecell[l]{co.cask.cdap:cdap-spark-core (1196), ch.cern.hadoop (114), com.facebook.presto.hadoop (247),\\ com.moz.fiji.mapreduce (381), com.clearspring.analytics (1650)} & \makecell[l]{com.h3xstream.findsecbugs (7), ch.cern.hadoop (114),\\ com.google.guave (16), org.slf4j (8), junit (73) } \\
    & \makecell[l]{org.apache.lucene\\ \\ }    & \makecell[l]{org.hibernate:search-analyzers (821), org.elasticsearch (231), nexus.bundles.elasticsearch (836),\\ org.jboss.seam.embedded (2451), org.apache.jackrabbit:oak-lucene (4710)} & \makecell[l]{org.jboss.seam.embedded (2451), junit (73), org.infinispan (581), \\org.apache.solr (948), org.elasticsearch (231)} \\
    & \makecell[l]{org.apache.opennlp\\ \\}   & \makecell[l]{com.graphaware.neo4j:nlp (10393), org.apache.uima (522), org.apache.tika (2003), \\org.apache.ctakes (3555),  edu.ucar:netcdf (2257)} & \makecell[l]{junit (73), java-nio-charset (13), java-util-regex (9),\\ org.apache.uima (522), org.slf4j (8)} \\
    & \makecell[l]{org.apache.maven\\ \\} & \makecell[l]{org.apache.maven.plugin-tools (120), org.codehaus.plexus (109), org.eclipse.aether (414), \\org.sonatype.plexus (851), org.apache.maven.shared (1274)
} & \makecell[l]{org.codehaus.plexus (109), org.apache.maven.plugin-tools (120), \\junit (73), org.eclipse.equinox (2774), commons-io (34) } \\
    \hline
    \multirow{4}{*}{\rotatebox{90}{\textbf{JS}}} & \makecell[l]{http} & \makecell[l]{https (31), url (14), express (3), request (17), querystring (32)}  & \makecell[l]{assert (10), https (31), url(14), express (3), net (48)} \\
    & \makecell[l]{mysql} & \makecell[l]{pg (299), redis (117), knex (343), mongodb (97), nodemailer (153)} & \makecell[l]{express (3), body-parser (13), async (25), lodash (12), request (17)} \\
    & \makecell[l]{jquery} & \makecell[l]{bootstrap (762), moment (27), underscore (22), angular (183), d3 (226)} & \makecell[l]{moment (27), datatables (520), underscore (22), backbone (272), lodash (12)}\\
    & \makecell[l]{gm (GraphicsMagick)} & \makecell[l]{imagemagick (1517), sharp (1040), connect-busboy (1913), jimp (1010), canvas (350)} & \makecell[l]{ async (25), request (17), express (3), lodash (12), crypto (16)} \\
    \hline
    \multirow{4}{*}{\rotatebox{90}{\textbf{Python}}} & \makecell[l]{scipy} & \makecell[l]{numpy (23), matplotlib (58), sklearn (123), mpl\_toolkits (219), pylab (228)} & \makecell[l]{numpy (23), matplotlib (58), warnings (32), sklearn (123), math (28)} \\
    & \makecell[l]{tensorflow} & \makecell[l]{keras (349), torch (476), cv2 (259), google (162), sklearn (123)} & \makecell[l]{numpy (23), time (8), object\_detection (8850), six (30), collections (11)} \\ 
    & \makecell[l]{http} & \makecell[l]{ssl (120), httplib (142), socketserver (276), socket (34), cookielib (376)} & \makecell[l]{urllib (19), httplib (142), io (18), urllib2 (65), urlparse (51)} \\
    & \makecell[l]{asyncio} & \makecell[l]{aiohttp (167), async\_timeout (574), uvloop (782), concurrent (147), websockets (625)} & \makecell[l]{logging (6), time (8), aiohttp (167), datetime (10), functools (15)} \\
  \hline
  \end{tabular}}
\end{table*}

\subsection{Measuring Predictive Power} \label{sec:evaluation-measure} 

To more objectively and quantitatively measure the quality of the embeddings, we
developed a variation of the contextual search algorithm that allows us to
interpret the search as a prediction task for which we can measure the accuracy
by using a held-out dataset (later called the validation set $D_v$) as ground truth data.

Our trained library embeddings describe a vector space $V_L$ with an embedding vector $v_l \in V_L$ for each library $l$ encountered in any project $p$ of our crawled training set $D_t$. 
In order to make verifiable predictions using these library vectors, we first calculate a project vector $v_p$ for each project $p$ in our training set $D_t$, by averaging all library vectors imported by $p$, i.e. the set $I_p$. 
This results in a new vector space $V_P$, called the project vector space:

$$
V_P = \{ v_p | p \in D_t, v_p = \sum_{l \in I_p}v_l/|I_p|  \}
$$

Similarly, given a set of context libraries $I_c$ for which we want to run a prediction, we calculate a context vector $v_c$ as

$$
v_c = \sum_{l \in I_c}v_l/|I_c|
$$

The algorithm then calculates the $k$ nearest neighbouring project vectors $P_{c,t}^k = \{v_p^i | i=1..k\}$ to the context vector $v_c$ from the training set $D_t$.  We now have $k$ projects that have a similar context of use.
It is therefore highly likely that these projects will include other libraries that are relevant for that context. We aggregate all libraries imported by any of the $k$ nearest neighbouring projects into an ordered set $I_c^k = \{ l_p \in I_p | p \in projects(P_{c,t}^k)\}$, sorted by the frequency count of each library.
The output of the algorithm are the top-$n$ most frequently occurring libraries in $I_c^k$, $n$ being the number of requested predictions. We refer to this output as $S^n$. 

To calculate the precision of predictions for a given context $I_c$, we 
calculate the subset of projects $P_{c,v}$ from the 
validation set $D_v$ that share this context: $P_{c,v} = \{ p | p \in D_v \wedge I_c \subset I_p \}$. Then, we aggregate all libraries imported by those 
projects into a set $I_{c,v} = \{ l_p | p \in P_{c,v} \wedge l_p \in I_p \}$. We
can then define the true positives $\text{TP} = (S^n \cap I_{c,v})$ and the false positives 
$\text{FP} = (S^n \setminus I_{c,v})$.
We then calculate a weighted precision $w(\text{TP}) / (w(\text{TP}) + |\text{FP}|)$ where each true positive $l_i \in \text{TP}$ is weighed by the fraction of context projects $p \in P_{c,v}$
that import $l_i$.
The TP are weighed to capture a degree of relevance, while the FP are not weighed, so that each import is either relevant to some degree, or totally irrelevant.


Running predictions ($n = 5$) on 1000 randomly sampled validation projects and averaging the precisions per project, we obtain the numbers listed in Table \ref{tab:precisionEval}.
Since the quality of the predictions highly depends on how semantically equivalent the closest neighbouring projects are to the given context, the precision on this task is taken to be a proxy measure for the quality of the trained embeddings.

\begin{table}[ht]
  \caption{Precision of evaluation algorithm. Left: considering only 1 NN project, Right: considering 20 NN projects} \label{tab:precisionEval}
  \centering
  \begin{tabular} { l | c c c | c c c }
    & \multicolumn{3}{c|}{$k = 1$ NN project} & \multicolumn{3}{c}{$k = 20$ NN projects} \\
    \hline
    context size & 2    & 5    & 10   & 2    & 5    & 10   \\ 
    \hline
    Java        & 0.84 & 0.80 & 0.70 & 0.85 & 0.71 & 0.49 \\ 
    JavaScript  & 0.75 & 0.56 & 0.47 & 0.71 & 0.45 & 0.32 \\ 
    Python      & 0.74 & 0.56 & 0.43 & 0.86 & 0.58 & 0.38 \\ 
  \end{tabular}
\end{table}

Has the model really learned something? In other words, what would be the precision if we were to 
use random library vectors rather than trained ones? Table~\ref{tab:randomEval} shows the results 
for the same sample of 1000 projects.
Comparing these results to those in Table~\ref{tab:precisionEval} reveals that trained vectors 
always lead to better predictions for all three ecosystems, indicating that the trained vectors 
encode information related to context.

\begin{table}[ht]
  \caption{Precision when using random library vectors} \label{tab:randomEval}
  \centering
  \begin{tabular} { l | c c c | c c c }
    & \multicolumn{3}{c|}{$k = 1$ NN project} & \multicolumn{3}{c}{$k = 20$ NN projects} \\
    \hline
    context size  & 2    & 5    & 10   & 2    & 5    & 10   \\ 
    \hline
    Java         & 0.55 & 0.56 & 0.56 & 0.58 & 0.60 & 0.41 \\ 
    JavaScript   & 0.45 & 0.37 & 0.37 & 0.55 & 0.40 & 0.26 \\
    Python       & 0.68 & 0.46 & 0.33 & 0.80 & 0.55 & 0.33 \\
  \end{tabular}
\end{table}

\subsection{Analogical Reasoning with Libraries}

Mikolov et al.~\cite{mikolov13distributed} describe how word embeddings can be used to
engage in analogical reasoning of the form $a$ is to $a^*$ as $b$ is to $b^*$.
To the best of our knowledge, there does not exist a standard dataset to describe
analogies among libraries. One way of constructing analogies for libraries
is to consider tools or frameworks that offer similar functionality (e.g. React and Vue
are both MVC frameworks for web applications), and to then consider more specialist
utility libraries within that ecosystem (e.g. react-bootstrap and bootstrap-vue
are both utility libraries that add support for Bootstrap HTML components to their
respective MVC framework).

In Table~\ref{tab:analogies} we list examples of such analogies for JavaScript.
Here $b^*$ is the to be predicted value and the ``pred rank'' column contains the rank of the predicted library $b^*$ among the nearest neighbours of $x^* = a^* - a + b$. 

Linzen~\cite{linzen2016issues} argues that in many cases analogical reasoning using vector embeddings is not based on an actual analogy as the correct result can often be predicted by just looking at the nearest neighbours of $b$ without considering the offset between $a^*$ and $a$. In many cases this offset will be small as $a^*$ and $a$ are themselves related and close to each other in the vector space. To quantify the sensitivity of our analogies to the offset between $a^*$ and $a$, the ``only-b rank'' column contains the rank of $b^*$ among the nearest neighbours of $b$. In all cases the ``pred rank'' is higher than the ``only-b'' rank, which provides evidence that adding the offset between $a^*$ and $a$ gets us closer to $b^*$.

\begin{table}[ht]
    \caption{Analogical Reasoning with Library Vectors} \label{tab:analogies}
    \centering
    \resizebox{\columnwidth}{!}{
    \begin{tabular} { c | c | c | c | c | c }
      \hline
      a                    & a*              & b                    & b* (prediction) & pred & only-b \\
      \textit{man}         & \textit{king}   & \textit{woman}       & \textit{queen}  & rank & rank   \\
      \hline
      express              & body-parser     & koa                  & koa-bodyparser  & 2    & 3      \\
      express              & express-session & koa                  & koa-session     & 4    & 12     \\
      connect-redis        & connect         & koa-redis            & koa             & 4    & 22     \\
      react-bootstrap      & react           & angular-ui-bootstrap & angular         & 1    & 7      \\
      react-bootstrap      & react           & bootstrap-vue        & vue             & 1    & 5      \\
      react-materialize    & react           & vue-mdl              & vue             & 1    & 369    \\
      angular-ui-bootstrap & angular         & reactstrap           & react           & 2    & 489    \\
      angular-ui-bootstrap & angular         & bootstrap-vue        & vue             & 1    & 5      \\
      mocha                & should          & sinon                & should-sinon    & 3    & 20     \\
      ts-loader            & webpack         & tsify                & browserify      & 6    & 31     \\
      webpack              & babel-loader    & browserify           & babelify        & 1    & 2      \\
      gulp                 & gulp-babel      & browserify           & babelify        & 1    & 2      \\
      axios-mock-adapter   & axios           & fetch-mock           & node-fetch      & 2    & 518    \\
      jquery               & axios           & request              & request-promise & 3    & 31     \\
      
    \end{tabular}}
\end{table}

\section{Related Work}

Our work fits into a larger body of work that approaches problems in software engineering by applying machine learning techniques to code and by using large open-source codebases as training data~\cite{hindle12naturalness, allamanis13mining, raychev15predicting, allamanis18survey}.

Alon et al.~\cite{alon2019code2vec} study how to build distributed representations of code snippets that capture deeper semantics of source code. Their training set is built by converting code into paths through the Abstract Syntax Tree (AST) of a code snippet. The embedding vectors are then used to predict method names for code snippets. In contrast, our work focusses on capturing the overall semantics of software projects or libraries as a whole, using only very shallow information about these projects and libraries (i.e. the import statements). Future work could focus on combining both library-level similarities with a deeper understanding of the code through code vectors
and more fine-grained definitions of library context of use, to further improve semantic code search.

Gude~\cite{python2vec} describes a method for learning word embeddings on Python source code tokens and uses the embeddings for recommending source code snippets to developers. We focus specifically on library imports to find similar and related libraries based on their context of use, by building specialized vector representations for libraries rather than generic word embeddings for source code tokens.

LeClair et al.~\cite{leclair18softclass} describe a neural model to automate software classification, by training embeddings based on both source code as well as textual project documentation. Rather than using internal library implementation details, we instead create import vectors for each library based on its context of use. Our objective is also to create library-specific import vectors that can be used for multiple downstream tasks, including contextual search or software classification tasks. Future work could combine both approaches, e.g. by leveraging our trained embeddings in their neural classification model for classifying software libraries.

Efstathiou et al.~\cite{efstathiou2018word} have applied word2vec to a corpus of 15GB of textual data drawn from Stack Overflow posts, in order to train word embeddings targeted at the software engineering domain. They show how training on a domain-specific corpus allows their embeddings to capture meaning of technical terms in a software engineering context that could not be captured when training on general-purpose text corpora. However, the textual data from Stack Overflow is still natural language text. Our work differs in that we apply word2vec to textual data extracted from source code.

Our work shows that word embedding models such as our adaptation of the skip-gram model lead to 
meaningful results when applied to library imports. There exist alternative word embedding models, 
such as word2vec continuous bag-of-words (CBOW)~\cite{mikolov13distributed}, 
FastText~\cite{DBLP:journals/corr/BojanowskiGJM16} (Skip-gram and CBOW), 
Doc2Vec~\cite{DBLP:journals/corr/LeM14} (DBOW and DM) and GloVe~\cite{pennington2014glove}.
A detailed comparison between these alternative models on our library similarity task is left
as future work. Initial comparisons between the word2vec skip-gram and CBOW models suggest that library vectors trained using skip-gram perform 5-10\% better on our evaluation task, but a more extensive study is needed to draw firm conclusions.



\section{Conclusion}

We adapted Mikolov et al.'s skip-gram model to train embeddings for libraries 
based on their context of use, by training on co-occurrence patterns of import statements
in code. Our experiments support the view that word embedding techniques can be applied
successfully not just to natural language but also to features extracted
directly from code.

To validate our work, we collected large-scale corpora of source code drawn from publicly 
available source code repositories and package managers. We contribute a detailed quantitative 
analysis of import co-occurrence patterns among six large library ecosystems and a detailed 
qualitative analysis of library vectors trained on three large library ecosystems. Our findings 
show that library vectors capture aspects of semantic similarity among libraries, clustering them 
according to specific domains or platform libraries. We demonstrate how library vectors can support 
applications such as contextual search and analogical reasoning.

More work is needed to objectively measure how well these vectors capture
semantic similarity of libraries. An 
open issue is the lack of a standard benchmark data set for this domain. For word embeddings, 
benchmarks exist such as the analogy dataset proposed by Mikolov et al.~\cite{mikolov13efficient} 
or word similarity tasks such as WordSim-353~\cite{finkelstein01placing}. Building an equivalent 
benchmark for library similarity would be highly valuable to push this line of work forward.

\bibliographystyle{ieeetran}
\bibliography{import2vec}

\begin{thebibliography}{10}
\providecommand{\url}[1]{#1}
\csname url@samestyle\endcsname
\providecommand{\newblock}{\relax}
\providecommand{\bibinfo}[2]{#2}
\providecommand{\BIBentrySTDinterwordspacing}{\spaceskip=0pt\relax}
\providecommand{\BIBentryALTinterwordstretchfactor}{4}
\providecommand{\BIBentryALTinterwordspacing}{\spaceskip=\fontdimen2\font plus
\BIBentryALTinterwordstretchfactor\fontdimen3\font minus
  \fontdimen4\font\relax}
\providecommand{\BIBforeignlanguage}[2]{{%
\expandafter\ifx\csname l@#1\endcsname\relax
\typeout{** WARNING: IEEEtran.bst: No hyphenation pattern has been}%
\typeout{** loaded for the language `#1'. Using the pattern for}%
\typeout{** the default language instead.}%
\else
\language=\csname l@#1\endcsname
\fi
#2}}
\providecommand{\BIBdecl}{\relax}
\BIBdecl

\bibitem{deshpande08total}
A.~Deshpande and D.~Riehle, ``The total growth of open source,'' in \emph{Open
  Source Development, Communities and Quality}, ser. IFIP – The International
  Federation for Information Processing, B.~Russo, E.~Damiani, S.~Hissam,
  B.~Lundell, and G.~Succi, Eds.\hskip 1em plus 0.5em minus 0.4em\relax
  Springer US, 2008, vol. 275, pp. 197--209.

\bibitem{mikolov13distributed}
T.~Mikolov, I.~Sutskever, K.~Chen, G.~S. Corrado, and J.~Dean, ``Distributed
  representations of words and phrases and their compositionality,'' in
  \emph{Advances in Neural Information Processing Systems 26}, C.~J.~C. Burges,
  L.~Bottou, M.~Welling, Z.~Ghahramani, and K.~Q. Weinberger, Eds.\hskip 1em
  plus 0.5em minus 0.4em\relax Curran Associates, Inc., 2013, pp. 3111--3119.

\bibitem{kim2014convolutional}
Y.~Kim, ``Convolutional neural networks for sentence classification,'' in
  \emph{Proceedings of the 2014 Conference on Empirical Methods in Natural
  Language Processing, {EMNLP} 2014, October 25-29, 2014, Doha, Qatar, {A}
  meeting of SIGDAT, a Special Interest Group of the {ACL}}, 2014, pp.
  1746--1751.

\bibitem{githubsearch}
\BIBentryALTinterwordspacing
G.~Inc. (2019) Advanced github search page. [Online]. Available:
  \url{https://github.com/search/advanced}
\BIBentrySTDinterwordspacing

\bibitem{lopes2017dejavu}
C.~V. Lopes, P.~Maj, P.~Martins, V.~Saini, D.~Yang, J.~Zitny, H.~Sajnani, and
  J.~Vitek, ``D{\'e}j\'{a}vu: A map of code duplicates on github,'' \emph{Proc.
  ACM Program. Lang.}, vol.~1, no. OOPSLA, pp. 84:1--84:28, Oct. 2017.

\bibitem{exampleproject}
\BIBentryALTinterwordspacing
J.-Y. Zhu, T.~Park, P.~Isola, and A.~A. Efros. (2019) Cyclegan and pix2pix in
  pytorch. [Online]. Available:
  \url{https://github.com/junyanz/pytorch-CycleGAN-and-pix2pix}
\BIBentrySTDinterwordspacing

\bibitem{zipf29freq}
G.~K. Zipf, ``Relative frequency as a determinant of phonetic change,''
  \emph{Harvard Studies in Classical Philology}, vol.~15, p. 1–95, 1929.

\bibitem{piantadosi14zipf}
S.~T~Piantadosi, ``Zipf's word frequency law in natural language: A critical
  review and future directions,'' \emph{Psychonomic bulletin \& review},
  vol.~21, 03 2014.

\bibitem{adamic2011zipf}
L.~A.~Adamic and B.~Huberman, ``Zipf's law and the internet,''
  \emph{Glottometrics}, vol.~3, 11 2001.

\bibitem{maaten2008visualizing}
L.~v.~d. Maaten and G.~Hinton, ``Visualizing data using t-sne,'' \emph{Journal
  of machine learning research}, vol.~9, no. Nov, pp. 2579--2605, 2008.

\bibitem{linzen2016issues}
T.~Linzen, ``Issues in evaluating semantic spaces using word analogies,''
  \emph{arXiv preprint arXiv:1606.07736}, 2016.

\bibitem{hindle12naturalness}
A.~Hindle, E.~T. Barr, Z.~Su, M.~Gabel, and P.~Devanbu, ``On the naturalness of
  software,'' in \emph{Proceedings of the 34th International Conference on
  Software Engineering}, ser. ICSE '12.\hskip 1em plus 0.5em minus 0.4em\relax
  Piscataway, NJ, USA: IEEE Press, 2012, pp. 837--847.

\bibitem{allamanis13mining}
M.~Allamanis and C.~Sutton, ``Mining source code repositories at massive scale
  using language modeling,'' in \emph{Proceedings of the 10th Working
  Conference on Mining Software Repositories}, ser. MSR '13.\hskip 1em plus
  0.5em minus 0.4em\relax Piscataway, NJ, USA: IEEE Press, 2013, pp. 207--216.

\bibitem{raychev15predicting}
V.~Raychev, M.~Vechev, and A.~Krause, ``Predicting program properties from "big
  code",'' in \emph{Proceedings of the 42Nd Annual ACM SIGPLAN-SIGACT Symposium
  on Principles of Programming Languages}, ser. POPL '15.\hskip 1em plus 0.5em
  minus 0.4em\relax New York, NY, USA: ACM, 2015, pp. 111--124.

\bibitem{allamanis18survey}
M.~Allamanis, E.~T. Barr, P.~Devanbu, and C.~Sutton, ``A survey of machine
  learning for big code and naturalness,'' \emph{ACM Comput. Surv.}, vol.~51,
  no.~4, pp. 81:1--81:37, Jul. 2018.

\bibitem{alon2019code2vec}
U.~Alon, M.~Zilberstein, O.~Levy, and E.~Yahav, ``code2vec: Learning
  distributed representations of code,'' \emph{Proceedings of the ACM on
  Programming Languages}, vol.~3, no. POPL, p.~40, 2019.

\bibitem{python2vec}
\BIBentryALTinterwordspacing
A.~Gude. (2016) Python2vec: Word embeddings for source code. [Online].
  Available:
  \url{https://gab41.lab41.org/python2vec-word-embeddings-for-source-code-3d14d030fe8f}
\BIBentrySTDinterwordspacing

\bibitem{leclair18softclass}
A.~LeClair, Z.~Eberhart, and C.~McMillan, ``Adapting neural text classification
  for improved software categorization,'' in \emph{2018 IEEE International
  Conference on Software Maintenance and Evolution (ICSME)}, Sep. 2018, pp.
  461--472.

\bibitem{efstathiou2018word}
V.~Efstathiou, C.~Chatzilenas, and D.~Spinellis, ``Word embeddings for the
  software engineering domain,'' in \emph{Proceedings of the 15th International
  Conference on Mining Software Repositories}, ser. MSR '18.\hskip 1em plus
  0.5em minus 0.4em\relax New York, NY, USA: ACM, 2018, pp. 38--41.

\bibitem{DBLP:journals/corr/BojanowskiGJM16}
\BIBentryALTinterwordspacing
P.~Bojanowski, E.~Grave, A.~Joulin, and T.~Mikolov, ``Enriching word vectors
  with subword information,'' \emph{CoRR}, vol. abs/1607.04606, 2016. [Online].
  Available: \url{http://arxiv.org/abs/1607.04606}
\BIBentrySTDinterwordspacing

\bibitem{DBLP:journals/corr/LeM14}
\BIBentryALTinterwordspacing
Q.~V. Le and T.~Mikolov, ``Distributed representations of sentences and
  documents,'' \emph{CoRR}, vol. abs/1405.4053, 2014. [Online]. Available:
  \url{http://arxiv.org/abs/1405.4053}
\BIBentrySTDinterwordspacing

\bibitem{pennington2014glove}
J.~Pennington, R.~Socher, and C.~D. Manning, ``Glove: Global vectors for word
  representation.'' in \emph{EMNLP}, vol.~14, 2014, pp. 1532--1543.

\bibitem{mikolov13efficient}
T.~Mikolov, K.~Chen, G.~Corrado, and J.~Dean, ``Efficient estimation of word
  representations in vector space,'' \emph{CoRR}, vol. abs/1301.3781, 2013.

\bibitem{finkelstein01placing}
L.~Finkelstein, E.~Gabrilovich, Y.~Matias, E.~Rivlin, Z.~Solan, G.~Wolfman, and
  E.~Ruppin, ``Placing search in context: The concept revisited,'' in
  \emph{Proceedings of the 10th International Conference on World Wide Web},
  ser. WWW '01.\hskip 1em plus 0.5em minus 0.4em\relax New York, NY, USA: ACM,
  2001, pp. 406--414.

\bibitem{ilspy}
\BIBentryALTinterwordspacing
S.~Pammer, D.~Grunwald \emph{et~al.} (2019) Icsharpcode ilspy .net decompiler.
  [Online]. Available: \url{http://www.ilspy.net/}
\BIBentrySTDinterwordspacing

\end{thebibliography}

\appendix

\textbf{Extracting imports from code} We provide more detail about how we extracted import directives from source files for the various languages under study.

As mentioned in Section~\ref{sec:importanalysis}, we analyze only (a subset of) the static imports.
For the purposes of this study it was not necessary to precisely extract every single import
dependency in code, but rather the most prevalent imports. To be able to extract import statements from the largest possible fraction of source files we used a lightweight extraction method, using combination of AST parsing and regular expression based parsing scripts.  
A summary of the raw imports statements extracted per language is shown in the table below: 

{\noindent
\begin{tabulary} {\textwidth}{ll}
 & \\
Lang. & Extracted import statements \\
\hline
Java & \texttt{import X.Y[.*];} \\
JS & \texttt{require('X')} \& \texttt{import Y[,..] from X} \\
Python & \texttt{import X.Y[.*]} \& \texttt{from X.Y import Z} \\
Ruby & \texttt{require[\_relative] `X'} \\
PHP & \texttt{use [function|const] X{\textbackslash}Y[{\textbackslash}\{Z,..\}]} \\
C\# & \texttt{using [static] X.Y[.*];} \\
\end{tabulary}
}

Below are language-specific import extraction notes:

\begin{enumerate}
    \item[Java] To extract source files for the MavenCentral JAR files, we restricted ourselves to projects that also uploaded the Java source files.
    \item[JS] For the NPM and GitHub JavaScript repositories, we ignored all files inside the \texttt{node\_modules} folder (contains source code of external library dependencies).
    \item[Ruby] We capture only the {\it require} statements, not the {\it load} statements that inject entire source files.
    \item[PHP] We capture only the {\it use} statements, not the {\it require} statements that inject entire source files.
    \item[C\#] To extract the source files for Nuget bundles, we used the DotNet ILSpy decompiler tool \cite{ilspy}.
\end{enumerate}

\end{document}